\documentclass[aps,prx,twocolumn,superscriptaddress,floatfix,nofootinbib,longbibliography]{revtex4-2}
\usepackage{graphicx,amsmath,amsfonts,amssymb,amsthm,xr}
\usepackage{epsfig,amsmath,amssymb,color,dsfont,upgreek,physics, bm}
\usepackage{mathrsfs}
\usepackage{mathtools}
\usepackage{bbold}
\usepackage{comment}
\usepackage{float}
\usepackage[bookmarks=true,colorlinks,linkcolor=Blue,urlcolor=Blue,citecolor=Blue]{hyperref}
\usepackage[dvipsnames]{xcolor}
\usepackage{orcidlink}

\renewcommand{\H}{\hat{\mathcal{H}}}
\newcommand{\hc}{\text{H.c.}}
\newcommand{\Zt}{$\mathbb{Z}_2$ }

\newcommand{\ad}{\hat{a}^{\dagger}}
\renewcommand{\a}{\hat{a}}

\newcommand{\tauZ}{\hat{\tau}^{z}}
\newcommand{\tauX}{\hat{\tau}^{x}}
\newcommand{\Sh}{\hat{S}}
\newcommand{\veci}{\bm{i}}
\newcommand{\vecj}{\bm{j}}

\renewcommand{\le}{\left(}
\renewcommand{\r}{\right)}

\begin{document}
\title{Exploring confinement transitions in \texorpdfstring{$\mathbb{Z}_2$}{Z2} lattice gauge theories\\ with dipolar atoms beyond one dimension}

\author{Matja\v{z} Kebri\v{c}${}{\orcidlink{0000-0003-2524-2834}}$}
\email{matjaz.kebric@colorado.edu}
\affiliation{JILA and Department of Physics, University of Colorado, Boulder, Colorado 80309, USA}
\affiliation{Department of Physics and Arnold Sommerfeld Center for Theoretical Physics (ASC), Ludwig-Maximilians-Universit\"at M\"unchen, Theresienstra\ss e 37, D-80333 M\"unchen, Germany}
\affiliation{Munich Center for Quantum Science and Technology (MCQST), Schellingstra\ss e 4, D-80799 M\"unchen, Germany}

\author{Lin Su${}{\orcidlink{0000-0002-0533-0311}}$}
\email{ls4211@columbia.edu}
\affiliation{Department of Physics, Harvard University, Cambridge, Massachusetts 02138, USA}
\affiliation{Department of Physics, Columbia University, New York, New York 10027, USA}

\author{Alexander Douglas${}{\orcidlink{0009-0001-2304-2292}}$}
\affiliation{Department of Physics, Harvard University, Cambridge, Massachusetts 02138, USA}

\author{Michal Szurek${}{\orcidlink{0000-0002-2683-341X}}$}
\affiliation{Department of Physics, Harvard University, Cambridge, Massachusetts 02138, USA}

\author{Ognjen Markovi\'{c}${}{\orcidlink{0000-0002-7094-0124}}$}
\affiliation{Department of Physics, Harvard University, Cambridge, Massachusetts 02138, USA}

\author{Ulrich Schollw\"ock${}{\orcidlink{0000-0002-2538-1802}}$}
\affiliation{Department of Physics and Arnold Sommerfeld Center for Theoretical Physics (ASC), Ludwig-Maximilians-Universit\"at M\"unchen, Theresienstra\ss e 37, D-80333 M\"unchen, Germany}
\affiliation{Munich Center for Quantum Science and Technology (MCQST), Schellingstra\ss e 4, D-80799 M\"unchen, Germany}

\author{Annabelle Bohrdt${}{\orcidlink{0000-0002-3339-5200}}$}
\affiliation{Department of Physics and Arnold Sommerfeld Center for Theoretical Physics (ASC), Ludwig-Maximilians-Universit\"at M\"unchen, Theresienstra\ss e 37, D-80333 M\"unchen, Germany}
\affiliation{Munich Center for Quantum Science and Technology (MCQST), Schellingstra\ss e 4, D-80799 M\"unchen, Germany}

\author{Markus Greiner${}{\orcidlink{0000-0002-2935-2363}}$}
\affiliation{Department of Physics, Harvard University, Cambridge, Massachusetts 02138, USA}

\author{Fabian Grusdt${}{\orcidlink{0000-0003-3531-8089}}$}
\email{fabian.grusdt@lmu.de}
\affiliation{Department of Physics and Arnold Sommerfeld Center for Theoretical Physics (ASC), Ludwig-Maximilians-Universit\"at M\"unchen, Theresienstra\ss e 37, D-80333 M\"unchen, Germany}
\affiliation{Munich Center for Quantum Science and Technology (MCQST), Schellingstra\ss e 4, D-80799 M\"unchen, Germany}

\begin{abstract}
Confinement of particles into bound states is a phenomenon spanning from high-energy to condensed matter physics, which can be studied in the framework of lattice gauge theories (LGTs).
Achieving a comprehensive understanding of confinement continues to pose a major challenge, in particular at finite matter density and in the presence of strong quantum fluctuations.
State-of-the-art quantum simulators constitute a promising platform to address this problem.
Here we study confinement in coupled chains of \Zt LGTs coupled to matter fields, that can be mapped to a mixed-dimensional (mixD) XXZ model.
We perform large-scale numerical matrix-product state calculations to obtain the phase diagram of this model, in which we uncover striped phases formed by the \Zt charges that can be melted at finite temperature or by increasing the tunneling rate.
To explore this setting experimentally, we use our quantum simulator constituted by erbium atoms with dipolar interactions in a quantum gas microscope, and observe the predicted melting of a stripe phase by increasing the particle tunneling rate.
Our explorative experimental studies of thermal deconfinement of \Zt charges motivate our further theoretical study of the mixD \Zt LGT, in which we predict a confined meson gas at finite temperature and low magnetization where thermal fluctuations destroy stripes but enable spontaneous commensurate spin order.
Overall, we demonstrate that our platform can be used to study confinement in \Zt LGTs coupled to matter fields, including long-range interactions and beyond one dimension, paving the way for future research of confinement in the quantum many-body regime.
\end{abstract}

\date{\today}
\maketitle

\section{Introduction}
Confinement of individual constituents into new emergent degrees of freedom plays an important role in many physical phenomena.
Lattice gauge theories (LGTs) are an invaluable tool to study confinement in both high-energy physics \cite{Wilson1974, Kogut1979, Fradkin1979, Drouffe1983, Greensite2020} and condensed matter settings \cite{Wegner1971, Kogut1979, Senthil2000, Wen2004, Sachdev2018}.
Despite impressive progress in recent decades, the confinement problem, where dynamical matter is coupled to a gauge field, is still not fully understood and remains an active field of research \cite{Greensite2003, Greensite2020}.
This is in part due to the complexity of numerical simulations of this problem, especially when the system dimension exceeds the simplest one-dimensional ($1+1$D) case \cite{Greensite2020, Grusdt2020, Gattringer2015, Gazit2017}.
For example, in Monte Carlo simulations at finite matter density, the sign problem for fermionic matter constitutes a significant limitation \cite{Grusdt2020, Gazit2017, Gattringer2015}.

Recently, quantum simulation experiments with ultracold atoms have achieved remarkable success in realizing Hubbard models and various spin systems \cite{Gross2017, Greiner2002, Bloch2008, Esslinger2010, Bohrdt2021}.
This also opened the possibility for quantum simulations of LGTs \cite{Schweizer2019, Barbiero2019, Goerg2019, Bender2018, Zohar2015, Homeier2023, HomeierPRB2021, Aidelsburger_2021, Banuls2020, Halimeh2025, Wiese2013, Mil2020, Kasper2017, Zache2018, Ott2021, Kasper2023, Hauke2013, Yang2016, Martinez2016, Farrell2024, Ciavarella2024, Bauer2023, Bauer2023qSimHEP, DiMeglio2024, Halimeh2025QsimOutEquil}.
Recent analog LGT implementations include Rydberg tweezer arrays via the Rydberg blockade mechanism \cite{Bernien2017, Surace2020, Semeghini2021}. A quantum link model has been implemented in an optical superlatice of bosons \cite{Yang2020, Zhou2022}.
Moreover, the study of confinement, often considered a hallmark phenomenon of LGTs, has recently become possible in programmable quantum simulators \cite{Cochran2024, Zhang2024, De2024, GonzalezCuadra2024, Cobos2025}.

Further advancement in quantum simulation of ultracold atoms with magnetic dipoles has opened up yet another avenue enabling the simulation of models containing long-range interactions \cite{Baier2016, Chomaz2022, Su2023}.
This allows for simulations of Hubbard models that contain interactions spanning beyond nearest-neighbor (NN) terms.
By considering hard-core bosons, one can map the problem onto a spin system with long-range interactions.
Such systems have been proposed to realize interesting phases of matter like spin liquids \cite{Bintz2024} and fractional Mott insulators or supersolids with hard-core bosons \cite{Sengupta2005}.

\begin{figure*}[t]
\centering
\epsfig{file=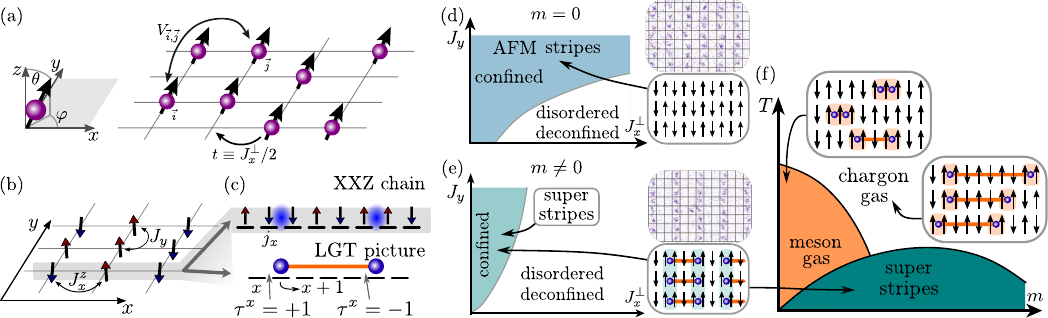}
\caption{Quantum simulation of a mixed-dimensional \Zt lattice gauge theory. (a) Sketch of our experimental system. We use external magnetic fields to tune the dipole-dipole interaction between strongly interacting erbium atoms hopping along the $x$-direction in the optical lattices. The presence (absence) of an atom can be mapped to an up-spin (down-spin) state on a given site.
(b) A sketch of the equivalent mixed-dimensional XXZ model Eq.~\eqref{eqMixedDXXZModel} in the spin basis, which we simulate numerically.
XXZ chains (in the $x$-direction) are coupled with an Ising interaction $J_y$ in the $y$-direction.
(c) Mapping between the XXZ chain and the \Zt lattice gauge theory coupled to matter, Eq.~\eqref{eq_mappedZ2LGT}. Orange lines denote the \Zt strings, where $\tau^{x} = -1$, and the absence of lines denotes anti-strings where the \Zt electric field takes the positive value, $\tau^{x} = +1$. AFM domain walls at the end of the strings corresponds to \Zt charges.
(d) Schematic ground state phase diagram for zero magnetization $m = 0$, constructed from our theoretical analysis, where the system forms a gapped AFM stripe-ordered state stabilized by the Ising interactions. We sketch spin orientation in the $z$-basis and show a typical experimental snapshot, where each occupied square correspond to individual erbium atoms.
(e) Sketch of the theoretical phase diagram at finite magnetization, where spins align into super-stripes stabilized by the inter-chain coupling $J_y$. We sketch the orientation of the spins and the corresponding \Zt charges. In addition we show an experimental snapshot of the super-stripes.
(f) Sketch of the theoretical phase diagram of coupled \Zt LGTs Eq.~\eqref{eq_mappedZ2LGT} as a function of temperature $T$ and magnetization $m$.
}
\label{FigOne}
\end{figure*}

In this work we explore confinement in coupled arrays of \Zt LGTs, realized in a mixed-dimensional (mixD) system.
Our starting point is a Bose-Hubbard model in the hard-core regime, with dipolar interactions extending across a two-dimensional plane, and hopping restricted to a single dimension.
This system can be modeled as a set of one-dimensional XXZ chains coupled together by an Ising interaction, giving rise to its mixD character.
In a next step, we map each XXZ chain to a \Zt LGT and show that sufficiently strong inter-chain coupling can lead to confinement.
In this mapping, density of the spin domain walls corresponds to a finite density of $\mathbb{Z}_2$-charged matter in the LGT; see Fig.~\ref{FigOne} (a)-(c) for an overview. 
We study the Bose-Hubbard experimentally in a quantum simulator, by trapping ultracold erbium atoms in an optical lattice \cite{Su2023, Phelps2020, Su2025} in a quantum gas microscope.
Our experimental results reveal various interesting phases of the coupled \Zt LGT chains, including a deconfined chargon gas and confined long-range ordered stripe phases, see Fig.~\ref{FigOne} (d)-(f) for a summary.

To gain a deeper theoretical understanding, we construct the phase diagram of the mixD XXZ model in the ground state and at finite temperature by employing matrix-product states (MPS) calculations \cite{Schollwoeck2011, Paeckel2019, hubig:_syten_toolk, hubig17:_symmet_protec_tensor_network}.
More precisely, we use the DMRG algorithm for ground state calculations \cite{Schollwoeck2011, White1992}, and the quantum purification scheme for the finite temperature calculations \cite{Zwolak2004, Feiguin2005, Feiguin2013, Nocera2016, Paeckel2019}.
In the ground state at zero net magnetization, we find antiferromagnetic AFM \textit{stripe} order, see Fig.~\ref{FigOne} (d), which is destroyed by quantum fluctuations when \Zt charge excitations proliferate.
At non-zero net magnetization, we find \textit{super-stripe} order where \Zt charges form extended line-like objects along $y$-direction, at regular distances along the $x$-direction, see Fig.~\ref{FigOne} (e).
These predictions are in qualitative agreement with our experimental findings.
In the language of the spin model, super-stripes feature incommensurate magnetism and closely resemble spin-charge stripes recently found in mixed-dimensional doped Hubbard models \cite{Bourgund2023}.
Our numerical calculations further predict the existence of a meson gas at finite temperature.
In this phase dynamical \Zt charges are confined into mesons in the \Zt LGT picture \cite{Grusdt2020} and commensurate magnetism is obtained in the spin model, which emerges at finite net magnetization as the super-stripe order is melted, see Fig.~\ref{FigOne} (f).
Finally, at temperatures high compared to the inter-chain coupling, the system becomes disordered, corresponding to the deconfined chargon gas phase of the \Zt LGT, which we also observe experimentally.
The meson gas proved to be least robust in our experiment, due to the effects of long-range interactions that stabilize the stripe phases.

This work is organized as follows: In Section~\ref{Model} we introduce the mixD XXZ model and present the mapping to coupled arrays of one-dimensional \Zt LGTs.
In addition, we discuss basic physical mechanisms of our models.
The following Section~\ref{ExperimentalResults} contains our experimental results, where we observe stripe phases that correspond to the confined phase in the \Zt LGT picture.
In Section~\ref{GroundStateNumerics} we further theoretically analyze the ground state phase diagram of a mixD XXZ model by performing large-scale numerical calculations.
Section~\ref{FiniteTnumerics} is dedicated to the theoretical study of the finite-temperature physics of the mixD XXZ model and the study of the confined meson gas.
We conclude in Section~\ref{Conclusion}.

\section{The mixed-dimensional XXZ model \label{Model}}
We start by introducing the two-dimensional (2D) Bose-Hubbard Hamiltonian that we can simulate by trapping
dipolar atoms in an optical lattice \cite{Su2023}
\begin{eqnarray}
    \H = - \sum_{\mu = x,y} \sum_{\langle \veci, \vecj \rangle_{\mu}} t_{\mu} \le \hat{b}^{\dagger}_{\veci} \hat{b}^{\phantom{\dagger}}_{\vecj} + \hc \r - \sum_{\veci<\vecj} V_{\veci,\vecj} \hat{n}_{\veci}^{b} \hat{n}_{\vecj}^{b} . 
    \label{eq_BoseHubbDipolar}
\end{eqnarray}
Here $\hat{b}^{\dagger}$ $(\hat{b})$ is the hard-core boson creation (annihilation) operator, $t_{\mu}$ is the NN tunneling amplitude and $V_{\veci,\vecj}$ denotes the dipolar interaction among hard-core bosons; see Fig~\ref{FigOne}(a).
Moreover $\langle \veci, \vecj \rangle_{\mu}$ labels nearest-neighbor pairs along the two spatial directions, $\mu = x, y$, and $\hat{n}^{b}_{\vecj} = \hat{b}^{\dagger}_{\vecj} \hat{b}^{\phantom{\dagger}}_{\vecj}$.
Dipolar interactions are defined as
\begin{equation}
    V_{\veci, \vecj}=V_0\frac{1-3((d_x/d)\sin\theta\cos\phi+(d_y/d)\sin\theta\sin\phi)^2}{d^3}, 
\end{equation}
where $\bm{d} = (d_x, d_y) \equiv \veci - \vecj$ and $d = |\veci - \vecj| $.
The polar and azimuthal angles, $(\theta,\phi)$, can be tuned by external magnetic fields \cite{Su2023}.
We first consider only nearest-neighbor interactions, i.e., pairs $\langle \veci, \vecj \rangle_\mu$ with $\veci-\vecj = \pm a \bm{e}_{\mu}$ where $a$ is the lattice spacing and $\bm{e}_{\mu}$ a unit vector along the $\mu = x,y$ direction, and ignore long-range interactions.
Although the interactions in our experiment are long-ranged, most of our theoretical study is focused on applying the above simplification to perform large-scale numerical calculations.
We will return to the effects of long-range tails when discussing our experimental results.

\subsection{Effective spin model}
Due to the hard-core property of the bosons, realized experimentally through strong on-site interactions \cite{Su2023}, the system can be mapped to a 2D spin-$1/2$ model, where occupied (vacant) lattice sites correspond to spin up (down) configurations in the $z$-basis.
Therefore, tunneling $t_\mu$ in Eq.~\eqref{eq_BoseHubbDipolar} realizes XX terms on bonds along $\mu = x,y$ in the effective Hamiltonian.
Parameters in Eq.~\eqref{eq_BoseHubbDipolar} can be tuned in a way that hopping of hard-core bosons only occurs in one spatial dimension: ${t_x \equiv t}$, ${t_y = 0}$.
Dipolar interactions $V_{\veci, \vecj}$ realize further-range Ising couplings in both lattice directions.
In the following we only consider the NN Ising coupling.
This yields a mixD XXZ model, i.e., a system of one-dimensional ($1$D) XXZ chains coupled by an Ising interaction; see Fig.~\ref{FigOne}(b).

The full Hamiltonian Eq.~\eqref{eq_BoseHubbDipolar} can thus be expressed as
\begin{eqnarray}
    \H = \H_{\rm XXZ} + \H_{r>1},
    \label{eqFullSpinModel}
\end{eqnarray}
where $\H_{r > 1}$ are the long-range spin interactions and $\H_{\rm XXZ}$ is the mixD XXZ model, defined as
\begin{multline}
    \H_{\rm XXZ} = \sum_{\langle \veci, \vecj \rangle_x } \left [ \frac{J_{x}^{\perp}}{2} \le \Sh^{+}_{\veci} \Sh^{-}_{\vecj} + \Sh^{+}_{\vecj} \Sh^{-}_{\veci} \r
    + J_{x}^{z} \Sh^{z}_{\veci} \Sh^{z}_{\vecj} \right ] \\
    - J_{y} \sum_{\langle \veci, \vecj \rangle_y} \Sh^{z}_{\veci} \Sh^{z}_{\vecj}.
    \label{eqMixedDXXZModel}
\end{multline}
Here we define $\Sh^{\pm}$ as the spin raising/lowering operator and $\Sh^{z}$ as the spin operator in the $z$ direction; $\langle \veci, \vecj \rangle_{x(y)}$ denotes NN bonds along $x(y)$-direction. The coupling $J_x^\perp = 2t$ is controlled by tunneling\footnote{We note that the prefactor in front of $J_x^\perp$ should be negative. However, we can preform a simple transformation ${\Sh^{x}_{\vecj} \rightarrow (-1)^{j_x} \Sh^{x}_{\vecj}}$, $ \Sh^{y}_{\vecj} \rightarrow (-1)^{j_x} \Sh^{y}_{\vecj} $, $\Sh^{z}_{\vecj} \rightarrow \Sh^{z}_{\vecj}$, which changes the sign.}, while $J_x^z = V_{\bm{e}_x, \bm{0}}$ and $J_y^z = V_{\bm{e}_y, \bm{0}}$ are controlled by the NN dipolar interaction strength that can be tuned independently.
We ignore the field terms $\propto V_{\veci,\vecj}\hat{S}^{z}_{\vecj}$, as we fix the filling in our system.
The density of the hard-core bosons, $n^{b}$, maps to magnetization of the spin model as $m = n^{b} - 1/2$.

In the following, we consider an anti-ferromagnetic (AFM) Ising interaction along the $x$-direction and a ferromagnetic (FM) Ising interaction along the $y$-direction, $J_x^{z}, J_y \geq 0$; see Fig.~\ref{FigOne}(b).
Since the spin raising/lowering operators only act in the $x$-direction we will refer to this model as the mixD XXZ model.
We note that the coupled XXZ model Eq.~\eqref{eqMixedDXXZModel}, has already been studied in the context of confinement in the regime of coupled XXZ ladders for zero magnetization, close to the ground state \cite{Lagnese2020, Lagnese2022}.
Such mixD spin systems are also often referred to as the \textit{Heisenberg-Ising} systems \cite{Wang2019, Lagnese2020, Lagnese2022}.
Here we study the system at finite magnetization and temperature, by coupling multiple XXZ chains.

Below we present results from MPS calculations of the mixD XXZ model Eq.~\eqref{eqMixedDXXZModel}, performed using \textsc{SyTen} \cite{hubig:_syten_toolk, hubig17:_symmet_protec_tensor_network}. 
If not stated otherwise, we consider open boundary conditions (OBC) in the $x$-direction and periodic boundary conditions (PBC) in the $y$-direction.
To be more precise, we consider a square lattice on a cylinder with length $L_x$ and circumference $L_y$, where $L_x \gg L_y$.

\subsection{Mapping to a lattice gauge theory \label{MappingToLGT}}
Next we show that the mixD spin model Eq.~\eqref{eqFullSpinModel} can be mapped to an array of coupled $1$D $\mathbb{Z}_2 $ lattice gauge theories (LGTs).
Within these $1$D chains, gauge mediated interactions can lead to confinement of individual \Zt charges (partons) into dimers (mesons) \cite{Kebric2021, Borla2020PRL}.
The main idea behind the mapping of the spin-$1/2$ model to the \Zt LGTs, is to consider the domain walls in the otherwise AFM ordered configuration as charges in the \Zt LGT; see Fig.~\ref{FigOne}(c).

We first introduce some notation: In each 1D chain along $x$-direction we introduce the dual site variables $x$, such that each site in the physical lattice $\vecj$, with $y$-component $j_y=y$, is characterized by the combination of a link $\langle x, x+1 \rangle$ on the dual lattice, and the chain index $y$, see Fig.~\ref{FigOne}(c). This allows us to define the \Zt electric fields on the links of the dual 1D lattice as $ \hat{\tau}^{x}_{\vecj} \equiv \hat{\tau}^{x}_{\langle x, x+1 \rangle, y}$  -- located across site $\vecj = (j_x, y)$ of the mixD lattice realized in our experiment. In addition, we define the \Zt gauge field, flipping $\hat{\tau}^{x}_{\langle x, x+1 \rangle, y}$, as $ \hat{\tau}^{z}_{\vecj} \equiv \hat{\tau}^{z}_{\langle x, x+1 \rangle, y}$.
\Zt electric and gauge fields are represented by Pauli $x$ and $z$ matrices on the dual links, respectively.
Finally, we define the \Zt charge number operators on the sites of the dual 1D lattice, labeled by the combination of $x$ and the chain index $y$ and corresponding to links $\langle \veci, \vecj \rangle_x$ oriented along $x$ in the physically realized lattice. I.e. $\hat{n}_{\langle \veci, \vecj \rangle_x} \equiv \hat{n}_{x,y} = \ad_{x,y} \a_{x,y}$, where $\ad_{x,y} (\a_{x,y})$ are hard-core boson creation (annihilation) operators.

Another key ingredient for the mapping to the \Zt LGT, is the one-dimensional Gauss law that can be expressed as a set of local operators \cite{Prosko2017}
\begin{equation}
    \hat{G}_{x, y} = \hat{\tau}^{x}_{\left \langle x-1, x \right \rangle , y} \hat{\tau}^{x}_{\left \langle x, x+1 \right \rangle , y} (-1)^{\hat{n}_{x, y}} .
    \label{eqGaussLawLGT}
\end{equation}
These operators are the generators of the \Zt gauge group and commute with the \Zt LGT Hamiltonian \cite{Prosko2017, Borla2020PRL}.
The eigenvalues $\hat{G}_{x,y} \ket{\psi} = g_{x,y} \ket{\psi}$ can take two values $g_{x,y} = \pm 1$.
We consider the physical sector, which is the sector without \Zt background charges where all eigenvalues are positive $g_{x,y} = + 1, \forall x,y$ \cite{Prosko2017, Kebric2021, Borla2020PRL}.
As a result the \Zt electric field changes its sign across a lattice site occupied by a \Zt charge, i.e., $\hat{G}_{x,y} \ket{\psi} = + \ket{\psi}$; see Fig.~\ref{FigOne}(c).
This allows us to introduce \Zt electric strings (anti-strings) that label the orientation of the \Zt electric fields $\tau^{x} = - 1$ ($\tau^{x} = +1$) \cite{Borla2020PRL, Kebric2021}.
Due to the Gauss law constraint, \Zt strings connect two individual partons; see Fig.~\ref{FigOne}(c).
Each parton pair is followed by an anti-string.
Moreover we can express the on-site \Zt charge number operator as
\begin{eqnarray}
    \hat{n}_{x,y} = \frac{1}{2} \le 1 -  \hat{\tau}^{x}_{\left \langle x-1, x \right \rangle ,y} \hat{\tau}^{x}_{\left \langle x, x+1 \right \rangle , y} \r .
    \label{eqGaussDenField}
\end{eqnarray}

Now we can perform the mapping from the mixD XXZ model to the mixD \Zt LGT.
We define the \Zt electric field on $\langle x, x+1 \rangle , y \equiv \vecj = (j_x, y)$ in terms of the spin operators as 
\begin{equation}
    \hat{\tau}^{x}_{\langle x, x+1 \rangle, y} \equiv \tauX_{\vecj} = 2 (-1)^{j_x} \Sh_{\vecj}^{z}.
    \label{eqTauXDef}
\end{equation}
Hence, by Eq.~\eqref{eqGaussDenField}, AFM domain walls of the spins $\Sh^{z}_{\vecj}$ correspond to \Zt charges.
With these identifications we can map the mixD spin model Eq.~\eqref{eqMixedDXXZModel} to the coupled arrays of 1D \Zt LGTs,
\begin{widetext}
\begin{eqnarray}
    \H_{\rm LGT} &=& t \sum_{x, y} \Big [ \le \ad_{x-1, y} \tauZ_{\langle x-1,x \rangle, y} \tauZ_{\langle x, x+1 \rangle, y} \a_{x+1,y} + \hc \r
    \le 1 -  \hat{n}_{x,y}  \r  \nonumber \\
    &+& \le \ad_{x-1,y} \tauZ_{\langle x-1, x \rangle, y} \tauZ_{\langle x, x+1 \rangle, y} \ad_{ x+1, y} + \hc \r
    \le 1 -  \hat{n}_{x, y} \r \Big ]
    + \frac{J_y}{4} \sum_{x,y} \tauX_{\langle x, x+1 \rangle, y+1} \tauX_{\langle x, x+1 \rangle, y}
    + \mu \sum_{x,y} \hat{n}_{x,y} + ...
    \label{eq_mappedZ2LGT}
\end{eqnarray}
\end{widetext}
Here, we define $t = t_x = \frac{J_x^{\perp}}{2}$, $\mu = \frac{J_x^{z}}{2}$, and ignore constant terms.
The term proportional to $\propto J_x^{\perp}$ thus maps to parton hopping and pair creation/annihilation processes along each chain.
The Ising interaction in the $x$-direction maps to the chemical potential, which controls the filling.
The last term ($...$) in Eq.\eqref{eq_mappedZ2LGT}, denotes the terms arising from long-range dipolar interactions, which are also gauge-invariant.
More details on the mapping can be found in Appendix \ref{AppendixLGTmapp}.
Since the Hamiltonian in Eq.~\eqref{eq_mappedZ2LGT} contains the pair creation and annihilation terms, the total parton number $\hat{N}^a = \sum_{x,y}\hat{n}_{x,y}$ is not conserved.

The Ising coupling between the chains, $\propto J_y$, leads to an effective string tension associated with the \Zt electric field.
On the mean-field level one can rewrite the inter-chain Ising interaction as
\begin{equation}
    \frac{J_y}{4} \sum_{x, y} \tauX_{\langle x, x+1 \rangle,y} \tauX_{\langle x, x+1 \rangle, y+1}
    \rightarrow h \sum_{x,y} \tauX_{\langle x, x+1 \rangle, y},
\end{equation}
where we define the mean-field \Zt electric field term as
\begin{equation}
    h = \frac{J_y}{4} \le \big \langle \tauX_{\langle x, x+1 \rangle, y+1} \big  \rangle + \big \langle \tauX_{\langle x, x+1 \rangle, y-1} \big  \rangle \r .
    \label{eqMFelectricfieldTerm}
\end{equation}
We also define the average expectation value of the $\hat{\tau}^{x}$ field in the chains $y\pm1$ as
\begin{equation}
    \langle \tauX_{\langle x, x+1 \rangle, y \pm 1} \rangle = 2 m_s^{y \pm 1} =  \frac{2}{L_x} \sum_{x} (-1)^j \langle \psi | \Sh^{z}_{(j_x, y \pm 1)} | \psi \rangle.
\end{equation}
This is the staggered magnetization of individual chains in the mixD XXZ basis, which we denote as $m_s^{y}$.
The coupling between the chains thus leads to a string tension $\propto h$, in the presence of commensurate magnetic order $\langle \tauX_{\langle x, x+1 \rangle, y \pm 1} \rangle \neq 0$.

We note that long-range dipolar interactions in Eq.~\eqref{eq_BoseHubbDipolar}, which we truncate to write the mixD XXZ model Eq.~\eqref{eqMixedDXXZModel}, do not break the Gauss law.
The effective LGT Hamiltonian contains additional gauge-invariant terms, and the language of \Zt LGT remains appropriate.
To be more precise, the long-range inter-chain coupling renormalizes the mean-field electric field term.
Furthermore, the long-range intra-chain AFM coupling, including on diagonal bonds, introduces spin frustration along the $x$-direction.
This results in a renormalized chemical potential term and beyond NN many-body charge interactions that do not change the physics of confinement, see Appendix~\ref{AppendixLGTmapp} for more details.
In the realistic experimental settings, the presence of disorder and finite temperature can also further modify the effective interactions.

\subsection{Basic physical mechanisms}\label{PhysMech}
We are now ready to discuss the basic physical mechanisms and simple limits of the mixD spin model, Eq.~\eqref{eqMixedDXXZModel}, and the corresponding coupled arrays of \Zt LGT, Eq.~\eqref{eq_mappedZ2LGT}.
\subsubsection{One-dimensional limit of the spin model}
In the decoupled regime, $J_y = 0$, the mixD model, Eq.~\eqref{eqMixedDXXZModel}, reduces to $L_y$ independent chains.
Each chain forms a gapless Luttinger liquid (LL) for generic parameter values $J_x^{\perp} / J_x^z$ and magnetization of the chain $m^{y} = \frac{1}{L_x} \sum_{j_x=1}^{L_x} S^{z}_{j_x, y} $ \cite{Giamarchi2004}.
The LL can be parameterized by the Luttinger parameter $K$, which can be extracted from the decay of the spin-spin correlations and characterizes the strength of interactions.
At zero magnetization, $m^{y} = 0$, the chains exhibit a BKT transition to a gapped antiferromagnetic phase for $J_x^{\perp} / J_x^z < 1$ \cite{Giamarchi2004}.

For weak coupling, $J_y \ll J_{x}^{\perp}, J_{x}^{z}$, the underlying physics remains quasi-one-dimensional.
The coupling between chains can be considered as additional interactions that renormalize the Hamiltonian of individual chains \cite{Giamarchi2004}.
The system remains a LL with renormalized interactions and modified Luttinger parameter.
With increasing couplings $J_x^{z}$ and $ J^{y}$ we eventually reach a 2D Ising limit, which we discuss next.

\subsubsection{Ising limit in two dimensions \label{BasicPhyIsing2D}}
In the limit $J_{x}^{\perp} \rightarrow 0$ the Ising couplings $J_x^{z}, J_y$ dominate the physics and the system reduces to a two-dimensional Ising model.
The FM coupling in the $y$-direction aligns the spins along $y$.
Conversely, along the $x$-direction, the AFM interactions induce long-range AFM order for zero magnetization, $m^y = 0$; see Fig.~\ref{FigOne}(d).
We dub this regime the AFM stripe phase, since the aligned spins in the $y$-direction form line-like objects with the period of two lattice sites in the $x$-direction.

For finite magnetization $|m^{y}| \neq 0$, the FM coupling between the chains also aligns the spins along the $y$-direction.
Along the $x$-direction, AFM interactions can stabilize incommensurate long-range magnetic order; see Fig.~\ref{FigOne}(e).
Line-like objects are formed by spins along the $y$-direction, similar to those in the $m= 0$ case.
For $|m^{y}| \neq 0$ these objects break the $\mathbb{Z}_M$ symmetry \cite{Bernien2017, Fendley2004}, where $M$ is associated with the period of such objects along the $x$-direction.
However, to accommodate the finite magnetization, the periodic spacing of such objects is larger than in AFM stripes and is related to magnetization $\propto 1 / 2 |m^{y}|$ \cite{Grusdt2020}.
We dub these objects \emph{super-stripes} to distinguish them from the AFM stripes with period two.

\subsubsection{One-dimensional \texorpdfstring{$\mathbb{Z}_2$}{Z2} LGT}
Next, we briefly review the physics of a one-dimensional \Zt LGT coupled to dynamical matter.
The \Zt gauge field is defined on links connecting neighboring lattice sites, and matter resides on the lattice sites \cite{Kebric2021, Borla2020PRL, Prosko2017},
\begin{equation}
    \H_{\rm 1D LGT} = -t \sum_{\langle i, j \rangle} \le \ad_j \tauZ_{\langle i, j \rangle} \a_i + \hc \r - h \sum_{\langle i, j \rangle} \tauX_{\langle i, j \rangle}.
    \label{eqZ2LGTOneDimDeff}
\end{equation}
Matter is minimally coupled to the \Zt gauge field $\hat{\tau}^{z}$ via the hopping term.
The corresponding \Zt electric field $\hat{\tau}^{x}$, which is its conjugate variable, induces dynamics in the gauge field \cite{Prosko2017}.

The corresponding Gauss law Eq.~\eqref{eqGaussLawLGT} introduces a set of local constraints on the Hilbert space.
We focus on the physical sector as defined in Sec.~\ref{MappingToLGT}. 
There, \Zt charges are connected in pairs with \Zt electric fields of the same sign, which we defined as the \Zt strings and anti-strings; see Fig.~\ref{FigOne}(c).

Finite value of the \Zt electric field term, $\propto h \tauX_{\langle x, x+1 \rangle}$, induces a linear confining potential associated with the \Zt electric string; see Fig.~\ref{FigOne}(c).
Hence, partons connected by the same string confine into mesons \cite{Borla2020PRL, Kebric2021, KebricNJP2023}.
It has been shown that any non-zero electric field term $h \neq 0$ results in confinement \cite{Borla2020PRL, Kebric2021, Kebric2026MF}, where partons belonging to the same meson cannot freely move away from each other.
This can be probed by calculating the \Zt gauge invariant Green's function that decays exponentially in the confined regime, and with a power law in the deconfined regime, $h = 0$ \cite{Borla2020PRL, Kebric2026MF}.
Mesons remain dynamical and form a LL when the Hamiltonian conserves the overall parton number \cite{Kebric2021, Borla2020PRL}.
Partons remain confined even at finite temperature $T$, where mesons can be clearly identified from string-length histograms obtained from snapshots \cite{Kebric2023FiniteT}.

Adding pair creation and annihilation terms,
\begin{eqnarray}
    \H_{\lambda} = \lambda \sum_{\langle i, j \rangle} \le \ad_j \tauZ_{\langle i, j \rangle} \ad_i + \hc \r,
\end{eqnarray}
to Eq.~\eqref{eqZ2LGTOneDimDeff} does not change the confinement of dynamical mesons in the one-dimensional \Zt LGT \cite{Kebric2026MF, Borla2021Kitaev}.
However, there is an additional confined state in the absence of the \Zt electric field term $h = 0$, for low and high parton fillings, corresponding to FM and AFM ordering of the \Zt electric field, respectively \cite{Kebric2026MF, Borla2021Kitaev}.

We note that in the deconfined regime, $h = 0$, gauge fields can be eliminated via Jordan-Wigner like transformation \cite{Kebric2026MF, Prosko2017}.
In this limit, we obtain a free parton model \cite{Kebric2026MF, Prosko2017}.
We also note that in a one-dimensional setting as well as in the mixD setting, the statistics of the charges is not important due to the Jordan-Wigner transformation, which can be used to map fermions to bosons and vice versa \cite{Kebric2021, Grusdt2020, Borla2020PRL}.

\subsubsection{Confinement in infinite arrays of coupled \texorpdfstring{$\mathbb{Z}_2$}{Z2} LGTs \label{BasicPhyisLGTconf}}
A similar mechanism as described for the one-dimensional \Zt LGT induces confinement in the mixD system, Eq.~\eqref{eq_mappedZ2LGT}.
Here, the confining field emerges due to the coupling of confined arrays of \Zt LGTs.
This can be understood on the mean-field level, where we require finite value of the mean-field \Zt electric field term Eq.~\eqref{eqMFelectricfieldTerm}, i.e., $h \neq 0$ is required to induce linear-confinement for the \Zt charges.
This is realized for finite coupling between XXZ chains $J_y >0$ and when the chains develop long-range AFM order, resulting in finite staggered magnetization $m_s^{y} \neq 0$, which occurs when $\langle \tauX_{\langle x, x+1 \rangle, y} \rangle \neq 0$.
Hence, a spontaneous commensurate magnetic order along the $x$-direction of the spin system induces finite value of the \Zt electric field $h \neq 0$ and confinement of partons into mesons.
Finite value of staggered magnetization develops spontaneously in confined chains with dynamical matter because \Zt strings tend to be on average much shorter than anti-strings in the confined regime \cite{Borla2020PRL, Kebric2023FiniteT}.
Hence, at any finite parton filling $n^y = \frac{1}{L_x} \sum_{x} \left \langle \hat{n}_{x,y} \right \rangle < 1$, the confined regime results in $\langle \tauX_{\langle x, x+1 \rangle, y \pm 1} \rangle > 0$.
The deconfined regime in Eq.~\eqref{eq_mappedZ2LGT} corresponds to the decoupled limit, $J_y = 0$, or in the absence of commensurate magnetic order, i.e., in the spin picture $m_s = 0$.

\begin{figure*}[t]
\centering
\epsfig{file=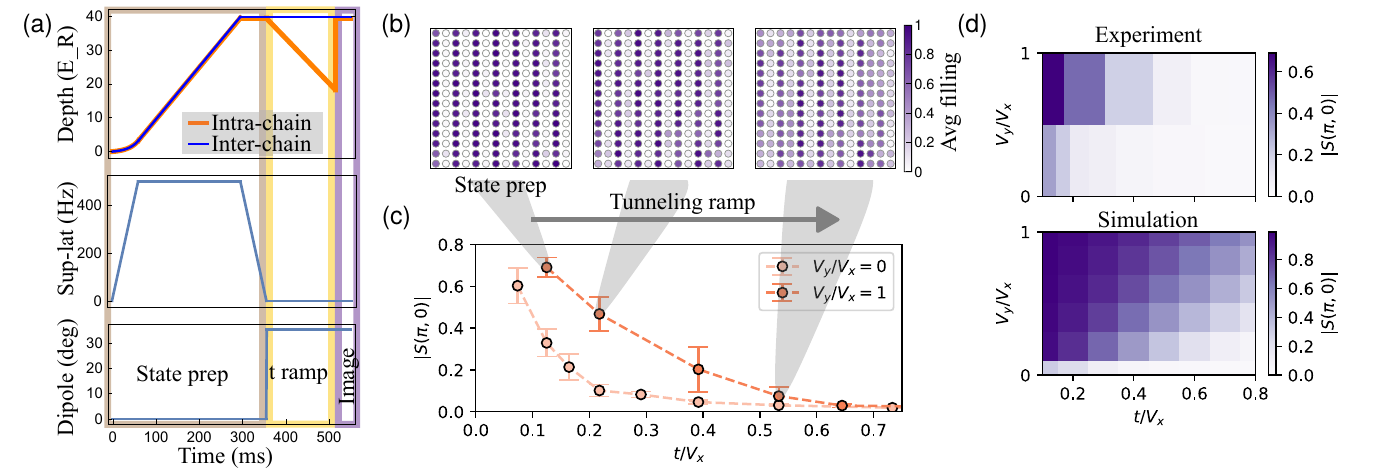, width=\textwidth}
\caption{Experimental simulation of the model Eq.~\eqref{eq_BoseHubbDipolar} at half filling, $n^{b} = 1/2$, in an optical lattice. (a) We first prepare low-entropy stripe or superstripe initial states with superlattice (brown box). Then we quench off the superlattice and ramp down tunneling slowly (yellow box), before finally performing site-resolved imaging (purple box). (b) At half filling ($m=0$), the prepared initial state has an average filling shown in the left most subfigure. We observe the gradual melting of the stripe as we turn up tunneling, as shown in the middle and right subfigure. (c) We evaluate the disconnected density-density correlator and take the Fourier transform to probe the structure factor corresponding to the stripe. We observe that the melting of the stripe takes place at lower tunneling energy when the inter-chain coupling is weak. (d) We compare the experimental results with DMRG ground state predictions, Eq.~\eqref{eqFourier2DSpinSpin} for $L_x = 40, L_y = 3$, and observe qualitative agreement.
}
\label{FigTwo_OneHalfFilling}
\end{figure*}

Next we consider the regime where $J_y \gg t$.
In the ground state, \Zt electric fields in different chains align along the $y$-direction to minimize the energy and they form line-like objects, regularly spaced in the $x$-direction.
The stripe phases of the mixD spin model Eq.~\eqref{eqMixedDXXZModel} are thus confined phases of the effective \Zt LGT Eq.~\eqref{eq_mappedZ2LGT}.
To be more concrete: AFM stripes correspond to a trivial vacuum state in the \Zt LGT basis because there are no domain walls in the AFM ordered spin configuration.
This corresponds to the absence of \Zt charges as governed by Eq.\eqref{eqGaussDenField}; see also Fig.~\ref{FigOne}(d).
At finite temperature $T>0$, \Zt charge excitations in AFM stripes remain confined due to strong staggered magnetization and finite inter-chain coupling, resulting in $h\neq 0$ in Eq.~\eqref{eqMFelectricfieldTerm}.
Super-stripes for $m \neq 0$ contain a finite \Zt charge density already in the ground state.
These \Zt charges are pinned to their corresponding stripes because the \Zt electric fields align along the $y$ direction to minimize their energy; see Fig.~\ref{FigOne}(e).

For low magnetization, \Zt excitations at finite temperature lead to melting of the stripe order.
However, due to strong interactions, the system develops a spontaneous AFM order in the mixD spin picture.
This mediates confinement in the \Zt LGT picture due to finite value of $h$, and \Zt charges confine into mesons, which remain mobile in the AFM background.
We coin this region the \emph{meson gas}, sketched in Fig.~\ref{FigOne}(f).
At higher temperature, such order is eventually destroyed and partons are deconfined.
For larger magnetization, spontaneous long-range AFM order is not possible and \Zt charges are deconfined also at lower temperature.
We dub this deconfined region the \emph{chargon gas}, see Fig.~\ref{FigOne}(f).

\section{Quantum simulation of lattice gauge theory using a dipolar quantum gas microscope \label{ExperimentalResults}}

We now describe our experimental realization of the lattice gauge theory, through the extended Bose-Hubbard model Eq.~\eqref{eq_BoseHubbDipolar}, using a tunable platform based on dipolar interactions.
Recent advances in using magnetic atoms in optical lattices have demonstrated the ability to adiabatically ramp systems into low-temperature many-body quantum phases with long-range, anisotropic dipolar interactions \cite{Su2023}.
These techniques allow us to prepare stripe and super-stripe initial states and study how the latter melt and \Zt charges deconfine.

\subsection{Experimental sequence}
We start our experiment from a Bose-Einstein Condensate (BEC) of $^{168}$Er prepared within 1 second~\cite{Phelps2020}.
The BEC is evaporated to the desired total atom number, so that we can simulate different magnetization $m$, as defined in the mixD model Eq.~\eqref{eqFullSpinModel}.
After compressing the BEC into a layer of our vertical lattice, we introduce an additional chemical potential term $\sum_{\vecj} \mu_S(-1)^{j_x}\hat{n}^b_{\vecj}$ in the Hamiltonian~\eqref{eq_BoseHubbDipolar}, where $\mu_S$ is the amplitude and $\vecj$ is the site index.
We experimentally realize this via our tunable-spacing, tunable-phase $1$D accordion lattice, which is aligned to one direction of the underlying square lattice, to serve as a superlattice \cite{Su2025, Su2025topological} so that low-entropy stripe and super-stripe patterns can be adiabatically prepared. Subsequently, we slowly turn on our tight-spacing, low-disorder 2D lattices \cite{Su2023} to ramp down tunneling $t$, as shown in the brown box of Fig.~\ref{FigTwo_OneHalfFilling}(a). 

After ramping to small tunneling $t\approx0.1V_x$ in roughly 200 milliseconds, we quench off the superlattice in 10 milliseconds to complete the preparation of the low-temperature initial states. We then quench the angle of the external magnetic field to change the anisotropy of the dipolar interaction and realize different ratios between $J_x^z$ and $J_y$. From here on, we simulate the Hamiltonian~\eqref{eq_BoseHubbDipolar}.
We then slowly increase $t$ exponentially to the desired value in roughly a hundred milliseconds (the orange box of Fig.~\ref{FigTwo_OneHalfFilling}(a)).
The choice of ramp duration balances the trade-off between minimizing non-adiabatic effects (favoring slower ramps) and reducing technical heating (which can be minimized by faster ramps).
At the end of the experiment, we quickly ramp up the lattices to quench tunneling in a millisecond, transfer to 2D accordion lattices, and expand the lattice spacing to a few microns for fast fluorescence imaging (the purple box of Fig.~\ref{FigTwo_OneHalfFilling}(a)).

\begin{figure*}[t]
\centering
\epsfig{file=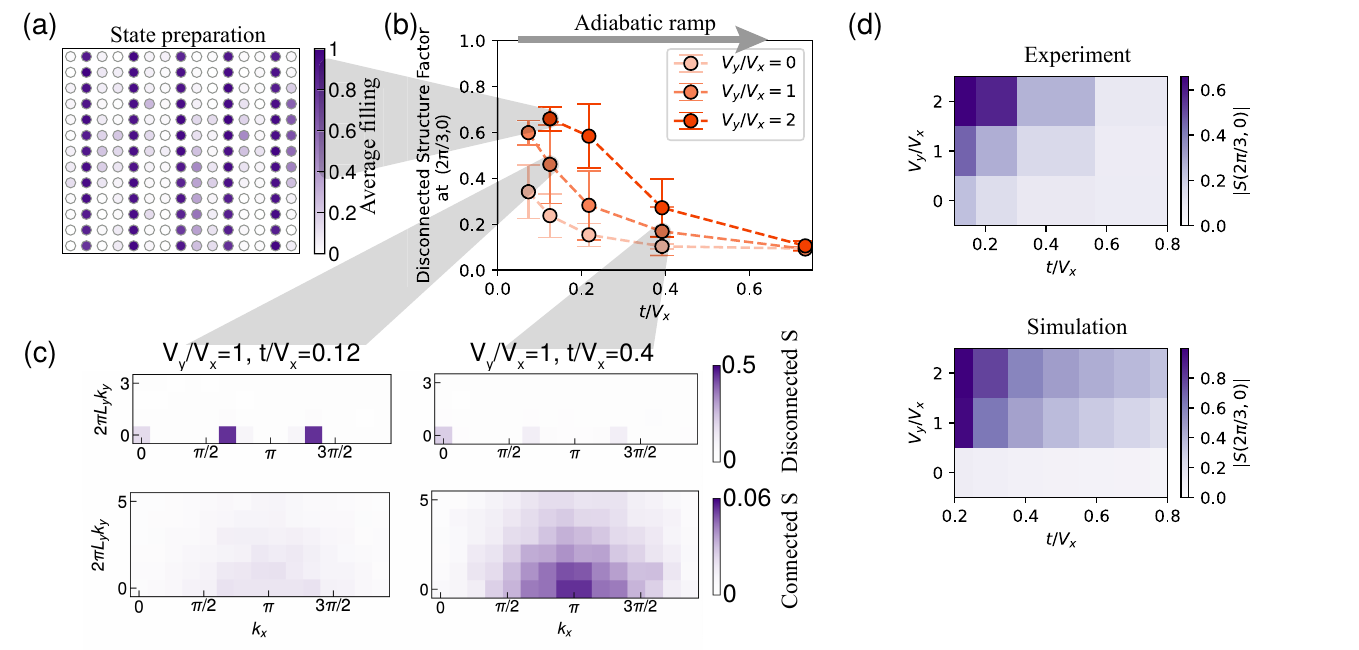, width=\textwidth}
\caption{Quantum simulation of the model in Eq.~\eqref{eq_BoseHubbDipolar}, with initial state preparation at one-third filling, $n^{b}=1/3$.
(a) We prepare the super-stripe initial state using tunable-spacing superlattices. (b) As we quench off the superlattices, we slowly turn up tunneling and observe the melting of the stripes. (c) Examples of the connected and disconnected structure factors. As tunneling is turned larger, the disconnected structure factor at ($2\pi/3$, 0) diminishes while the connected structure factor near ($\pi$, 0) intensifies. (d) We compare the experimental data with the DMRG ground state simulated results, Eq.~\eqref{eqFourier2DSpinSpin} for $L_x = 36, L_y = 3$, and observe qualitative agreement.
}
\label{FigThree_OneThirdFilling}
\end{figure*}

\subsection{Correlation functions}
We calculate the disconnected density-density correlator
\begin{equation}
\label{eq:d_correlation}
D_\mathbf{d}=\frac{4}{N_{\mathbf{d}}}\sum_{\mathbf{d}=\mathbf{d}_{i,j}}\langle \hat{n}_{\veci}^{b} \hat{n}_{\vecj}^{b} \rangle,
\end{equation}
and the connected density-density correlator
\begin{equation}
\label{eq:c_correlation}
C_\mathbf{d}=\frac{4}{N_{\mathbf{d}}}\sum_{\mathbf{d}=\mathbf{d}_{i,j}}(\langle \hat{n}_{\veci}^{b} \hat{n}_{\vecj}^{b} \rangle-\langle \hat{n}_{\veci}^{b}\rangle\langle \hat{n}_{\vecj}^{b}\rangle),
\end{equation}
where the sum runs over $N_{\mathbf{d}}$ pairs of lattice sites separated by the distance vector $\mathbf{d}$, and averages are taken over hundreds of experimental runs. From these correlators, we compute the structure factors, defined as \cite{Su2023}
\begin{equation}
\label{eq:d_structure_factor}
S_D(\mathrm{q})\propto\sum_{\mathbf{d}}e^{i\mathbf{q}\cdot\mathbf{d}}D_\mathbf{d},
\end{equation}
and
\begin{equation}
\label{eq:c_structure_factor}
S_C(\mathrm{q})\propto\sum_{\mathbf{d}}e^{i\mathbf{q}\cdot\mathbf{d}}C_\mathbf{d}.
\end{equation}

\subsubsection{Zero magnetization: \texorpdfstring{$\mathbb{Z}_2$}{Z2} vacuum}
First, we realize the system with zero magnetization ($m=0)$, which maps to half-filling in the Bose-Hubbard model.
We prepare initial states as shown in Fig.~\ref{FigTwo_OneHalfFilling}(b) left.
Then we slowly ramp up tunneling and measure the disconnected structure factor at $\mathbf{k} =  (\pi, 0)$, revealing long-range AFM correlations in the system, see Fig.~\ref{FigTwo_OneHalfFilling}(c).
These correlations directly correspond to the AFM stripes, where spins align along the $y$-direction and form an AFM pattern, with the period of two lattice sites, along the $x$-direction.
Considering the height of the structure factor peak at $(\pi, 0)$, shown in Fig.~\ref{FigTwo_OneHalfFilling}(c), we observe that the AFM stripes for $V_y = V_x$ melt at significantly higher tunneling $t$, compared to vanishing of the AFM correlations in the 1D limit, $V_y = 0$.

AFM stripes correspond to the vacuum state of the coupled arrays of \Zt LGTs, where the \Zt charge excitations are confined due to strong inter-chain coupling, as discussed in Sec.~\ref{BasicPhyisLGTconf}.
Our experimental results show that for finite inter-chain coupling $V_x = V_y$, increasing tunneling $t = \frac{1}{2} J_{x}^{\perp} \gtrsim 0.5 V_x = 0.5 J_x^{z}$, results in melting of the AFM stripes into a disordered state where commensurate magnetic order is destroyed, resulting in deconfined \Zt charges.
Plotting the results of the structure factor in 2D as a function of $V_x$ and $V_y$, we see qualitative agreement with the DMRG simulations, where we simulate the mixD XXZ model Eq.~\eqref{eqMixedDXXZModel}; see Fig.~\ref{FigTwo_OneHalfFilling}(d).
The structure factor decreases significantly faster in the decoupled regime $V_y = 0$ compared to the coupled regime $V_y = V_x$, in agreement with our theoretical prediction that the inter-chain coupling stabilizes the confined phase with long-range spin order; see Sec.~\ref{BasicPhyIsing2D}.
We discuss our DMRG simulations in detail in Sec~\ref{GroundStateNumerics}.

\begin{figure*}[t]
\centering
\epsfig{file=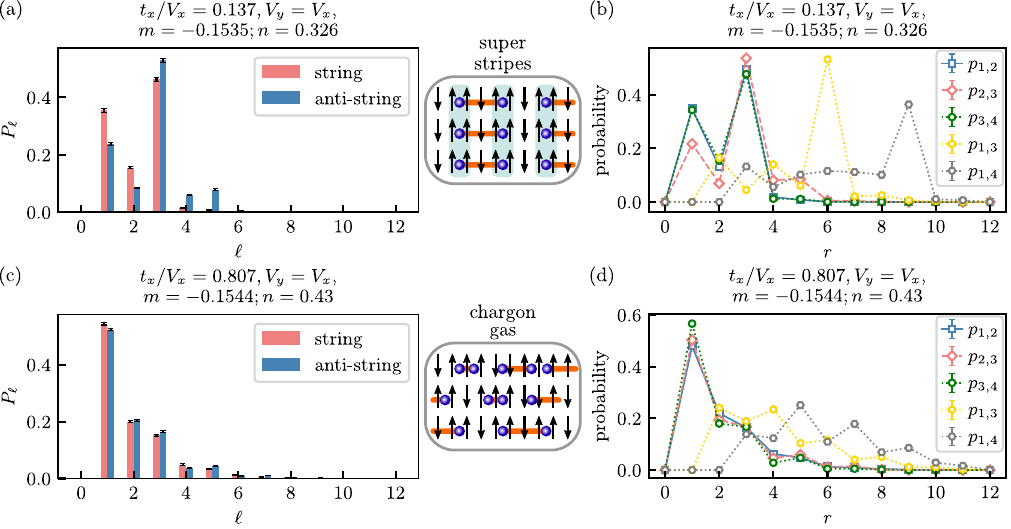}
\caption{Experimental histograms of parton spacing distributions at finite inter-chain coupling $V_y = V_x$ and magnetization $|m| \approx 0.15$.
(a) String and anti-string length distributions, as defined in Eq.~\eqref{eqTauXDef}, for low tunneling rate $t_x/V_x = 0.137$, have qualitatively similar shapes and peaks sharply at the (anti-)string length $\ell = 3$.
(b) Probability distributions of the distance between different partons $p_{a, b}(r)$, where $r$ is the distance between $a$-th and $b$-th parton from the left corner of the snapshot.
For a low tunneling rate $t_x/V_x = 0.137$, the distributions $p_{1,2}, p_{2,3}, p_{3,4}$ have a peak at distance $r = 3$, and do not quantitatively differ from each other, which establishes the super-stripe phase.
(c) For a higher tunneling rate $t_x/V_x = 0.807$, string and anti-string length distributions have a maximum at the shortest possible value $\ell = 1$ and have qualitatively equal shapes, which we interpret as a distinct feature of the chargon gas phase.
(d) Probability distributions $p_{a,b}(r)$, for the same parameters as in (c) reveal a peak at $r = 1$ for $p_{1,2}, p_{2,3}, p_{3,4}$, which establishes a chargon gas regime.
The errors are estimated via the bootstrap method.
}
\label{FigFour_ExperimentStrings}
\end{figure*}

\subsubsection{Finite magnetization: \texorpdfstring{$\mathbb{Z}_2$}{Z2} charged case}
Next, we study the model for non-zero magnetization in the Bose-Hubbard picture.
We set the spacing of our superlattice using a galvanometer \cite{Su2023} to prepare an initial state as shown in Fig.~\ref{FigThree_OneThirdFilling}(a) with $m\approx-0.167$, i.e., at one-third filling, $n^{b} = 1/3$.
Then we slowly ramp up the tunneling (see detailed discussion in Appendix~\ref{ExperimentalDetailsAppendix}) and probe the structure factor as shown in Fig.~\ref{FigThree_OneThirdFilling}(b).
We observe peaks in the disconnected structure factor at $\bm{k} = (2 \pi/ 3, 0)$, which correspond to the translational symmetry breaking, super-stripe phase.
Furthermore, structure factor peaks directly relate to the magnetization of the system, $k_x = (1 \pm 2|m|) \pi = \frac{2}{3} \pi, \frac{4}{3} \pi$; see Fig.~\ref{FigThree_OneThirdFilling}(c).

The super-stripes correspond to the confined, striped phase of \Zt charges with a period of three lattice sites, see Fig.~\ref{FigOne}(e).
The structure factor peak decays with increasing tunneling rate $t$, which indicates the melting of the super-stripe order.
The decay of the structure factor is significantly decreased with increasing inter-chain coupling $V_y$, which agrees with our theoretical predictions: the 2D Ising limit corresponds to a perfect super-stripe order, which is destroyed by increasing the XX term strength in the mixD Hamiltonian Eq.~\eqref{eqMixedDXXZModel}.
The observed disconnected structure factor qualitatively agrees with the DMRG ground state simulation for the mixD Hamiltonian Eq.~\eqref{eqMixedDXXZModel}, as shown in Fig.~\ref{FigThree_OneThirdFilling}(d).

Moreover, the connected structure factor starts to develop a peak near $\bm{k} = (\pi,0)$ as shown in Fig.~\ref{FigThree_OneThirdFilling}(c) bottom, as we ramp the tunneling amplitude further up.
This indicates the formation of the commensurate AFM order, suggesting the emergence of a meson gas phase.

\subsection{Microscopic analysis of the super-stripes melting}
We analyze the melting of the super-stripes further by studying the distances between the spinons (i.e., AFM domain walls) along the $x$-direction in the AFM background.
This corresponds to the spacing between the \Zt charges in the \Zt LGT picture.
We first analyze the string and anti-string lengths $\ell$, see Fig.~\ref{FigOne}(c). 
String and anti-string lengths correspond to the spacing between the partons connected by the \Zt electric fields of the same orientation \cite{Kebric2021}, as defined in Eq.~\eqref{eqTauXDef}.
We define strings (anti-strings) as a consecutive sequences of the \Zt electric fields with $\left \langle \tauX \right \rangle = -1$ $(\left \langle \tauX \right \rangle = +1)$.

We also study the probability distributions of the parton distances, $p_{a, b}(r)$ \cite{Grusdt2020}.
In each snapshot, we determine the position of each parton and label them as they appear in the chain from left to right.
We record the distance $r$ between $a$-th and $b$-th parton starting from the left and plot the probability distributions $p_{a, b}(r)$.

The identity of strings and anti-strings cannot be easily determined in an individual snapshot, meaning that the distance between $a$-th and $(a+1)$-th parton can be either a string or an anti-string in the \Zt picture; see Fig.~\ref{FigOne}.
For that reason we assign the identity of the strings and anti-strings based on the total staggered magnetization, which we discuss in Section~\ref{SectionV_strings}.
We perform the same procedure for the probability distributions $p_{a, b}(r)$, see Appendix~\ref{AppendixPartonSpacing} for further details.

We consider finite magnetization regime $|m|\approx0.167$, where super-stripes are stabilized at low temperature.

\subsubsection{Super-stripes}
We start our analysis at one-third filling, $n^{b} = 1/3$ ($|m| = 0.167$), in the super-stripe regime.
For a low tunneling rate $t_x/V_x = 0.137$ and for finite inter-chain coupling $V_y = V_x$, we observe the highest probability peak $P_{\ell}$ in our histograms for string and anti-string length $\ell = 3$, see Fig.~\ref{FigFour_ExperimentStrings}(a).
Moreover we only observe a small qualitative difference between the strings and anti-strings, which we attribute to the choice of determining the identity of strings and anti-strings.
This indicates a super-stripe phase with a period of three lattice sites as sketched in Fig.~\ref{FigFour_ExperimentStrings}.
For details on the construction of histograms from the snapshots see Section~\ref{SectionV_strings}.

For magnetization $|m|\approx 0.167$ and low tunneling rate $t_x/V_x = 0.137$, we observe that consecutive partons have the highest probability of being three lattice sites away from each other $p_{a, a+1} (r = 3) > p_{a,a+1}(r\neq 0) $, see Fig.~\ref{FigFour_ExperimentStrings}(b).
Moreover, these distributions are qualitatively similar to each other, $p_{1,2}(r) = p_{2,3}(r) = p_{3,4}(r)$.
We also find that the distance probabilities between $a$-th and $(a+2)$-th parton peak at $r = 6$ and that $p_{1,4}(r)$ has a peak at $r = 9$.
These are the signatures of the super-stripe regime with a period of three lattice sites, as sketched in Fig.~\ref{FigFour_ExperimentStrings}(a).
It is also consistent with the measured parton filling $n = 0.326$, which is close to the filling of $n^a = 1/3$ expected for the super-stripe regime with the period of three lattice sites.
For details on construction of the density probabilities, see Appendix~\ref{AppendixPartonSpacing}.

\subsubsection{Deconfined chargon gas}
At the same magnetization $|m| \approx 0.167$, but for higher tunneling rate $t_x/V_x = 0.807$, the string and anti-string length histograms peak at $\ell = 1$, and gradually decay for larger values of $\ell$, see Fig.~\ref{FigFour_ExperimentStrings}(c).
Parton separation distributions for the same parameter regime are shown in Fig.~\ref{FigFour_ExperimentStrings}(d).
Consecutive partons have the highest probability of being located only one site away, and the distributions $p_{1,2}(r)$, $p_{2,3}(r)$, $p_{3,4}(r)$ are qualitatively similar.
The distributions $p_{1,3}$ and $p_{1,4}$ also change qualitatively, in particular the peaks at $p_{1,3}(r = 6)$ and $p_{1,3}(r = 9)$ disappear and the distributions become much broader compared to the distributions for $t_x/V_x = 0.137$.
These results indicate the chargon gas of deconfined \Zt partons, where partons are able to move freely along the $x$-direction; see the sketch in Fig.~\ref{FigFour_ExperimentStrings}.

\subsubsection{Signatures of the meson gas}
\begin{figure}[t]
\centering
\epsfig{file=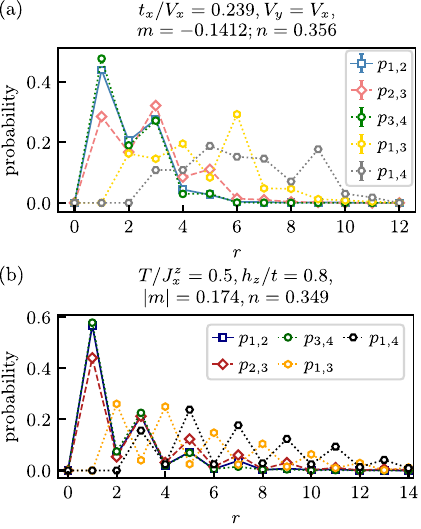}
\caption{
Probability distribution of the distance between partons, $p_{a,b}(r)$.
(a) Experimental results for tunneling amplitude $t_x / V_x = 0.239$, and finite magnetization $|m| = 0.141$, exhibit qualitatively different distributions for $p_{2s-1,2s}(r)$ and $p_{2s,2s+1}(r)$. In particular peaks at $r = 1$ are higher for $p_{1,2}, p_{3,4}$ than for $p_{2,3}$.
(b) Similarly, numerical MPS results (discussed in Section~\ref{FiniteTnumerics}), show a qualitatively different behavior of $p_{1,2}(r) \approx  p_{3,4}(r)$ from $p_{2,3}(r)$.
Numerical calculations were performed on a cylinder of length $L_x = 20$ and circumference $L_y = 3$ for parameters $J_{x}^{\perp} / J_{x}^{z}= 0.4$, $J_{y} / J_{x}^{z} = 0.8$, which corresponds to $t = 0.2V_{x}$ and $V_y = 0.8 V_{x}$, comparable to the experimental values.
Furthermore, the magnetization and the parton filling in the experimental results and numerical calculations, are comparable.
}
\label{FigFive_MesonGasDistances}
\end{figure}

A meson gas phase should exhibit a dramatic difference between the probabilities $p_{2s-1,2s}(r)$ and $p_{2s,2s+1}(r)$, where we assume that a string connects partons $2s-1$ and $2s$ and an anti-string connects partons $2s$ and $2s+1$.
Hence, the $p_{2s-1,2s}(r)$ should exhibit a narrow peak at a short distance $r \approx 1$, and $p_{2s,2s+1}(r)$ should have a broad distribution with a peak $r \gg 1$ for low filling \cite{Grusdt2020}.
Our experimental results at finite magnetization $|m| \approx 0.167$, for intermediate tunneling rate $t_x/V_x = 0.239$, exhibit a qualitative difference in peaks $p_{1,2}(r = 1) > p_{2,3}(r = 1)$, which we interpret as a signature of a meson gas phase; see Fig.~\ref{FigFive_MesonGasDistances}(a).
These results qualitatively agree with the numerical MPS simulations for $p_{a, a+1}(r = 1)$ at comparable magnetization $|m| \approx 0.17$ and parameter regime in Fig.~\ref{FigFive_MesonGasDistances}(b).
However, the qualitative agreement of the experimental results with the numerical results becomes worse for $r>1$, which prevents us to conclusively confirm the existence of a confined meson gas in our experiment.
In particular, the experimental results still show high probabilities at distances $r = 3$ for the NN spacings.
This could suggest that we still see some signatures of the super-stripe order after the time evolution in the experiment, which may be stabilized by long-range interactions, see Appendix~\ref{AppendixLGTmapp} and ~\ref{ExperimentalDetailsAppendix}.
Our numerical calculations, performed for comparable parameter values $J_{x}^{\perp} / J_{x}^{z}= 0.4$, $J_{y} / J_{x}^{z} = 0.8$, yield approximately similar magnetization $m=0.174$ (experiment $|m| \approx 0.141$) and similar average parton filling $n^{a} \approx 0.35$ for temperature $T = 0.5V_x$.

\begin{figure*}[t]
\centering
\epsfig{file=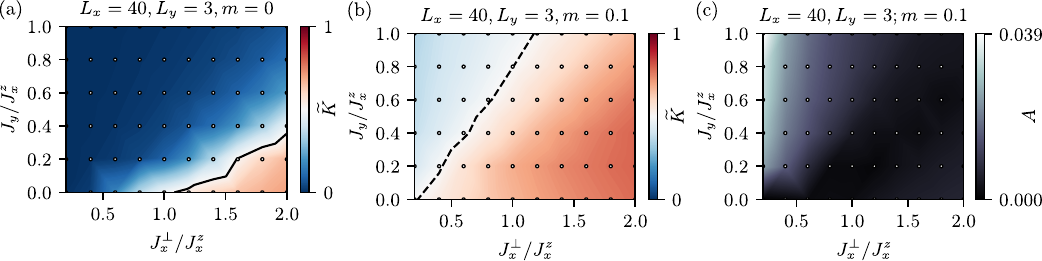}
\caption{
Numerical study of spin-spin correlations $\langle \Sh^{z}_{j} \Sh^{z}_{j+x} \rangle$ in the mixD XXZ model, Eq~\eqref{eqMixedDXXZModel}. (a) LL-like parameter $\tilde{K}$ obtained by fitting the DMRG data with Eq.~\eqref{eqLLCorrelationsZmag} in the $x$-direction for zero magnetization $m = 0$.
The system forms an ordered long-ranged AFM state for $J_{x}^{\perp} / J_x^{z} < 1$, resulting in low values of $\tilde{K} < 1/2$.
This state is extended for non-zero inter-chain coupling $J_y > 0$ also for larger $J_{x}^{\perp} / J_x^{z} > 1$.
(b) For finite magnetization $m = 0.1$, the system features an increasing LL-like parameter $\tilde{K}$ with increasing value of $J_x^{\perp}$ and $J_y$, which indicates enhanced spin-spin correlations.
Black lines mark the value of the LL-like parameter $\tilde{K} = 1/2$.
(c) Parameter $A$ in Eq.~\eqref{eqStripeFct}, which saturates to a finite value in the long-range limit when the system forms super-stripes.
The system size in all cases is $L_x = 40$ and $L_y = 3$ with PBC in the $y$-direction.
}
\label{FigSix_correlations}
\end{figure*}

We discuss the meson gas in further detail theoretically in Section~\ref{FiniteTnumerics}, where we also provide details on our numerical calculations using MPS.

\section{Ground state phase diagram \label{GroundStateNumerics}}
In this section we provide a deeper theoretical understanding of the ordered stripe and super-stripe phases observed experimentally.
To this end we analyze the ground state phase diagram of the mixD XXZ model Eq.~\eqref{eqMixedDXXZModel}.

The physics of the mixD XXZ model depends on the magnetization of the system and we separately study the model at zero and finite magnetization.
We first consider the spin-spin correlations along the $x$-direction and relate our results in the one-dimensional limit to the LL theory.
We find enhanced AFM correlations due to inter-chain coupling that indicate the AFM stripe phase, see Fig.~\ref{FigOne}(d).
Furthermore, we study super-stripes for $m \neq 0$, by considering the Fourier transformation of the full spin-spin correlations, which remain stable for low values of the XX term.

We use DMRG \cite{Schollwoeck2011, White1992} to simulate the mixD model Eq.~\eqref{eqMixedDXXZModel}, on a cylinder using \textsc{SyTen} \cite{hubig:_syten_toolk, hubig17:_symmet_protec_tensor_network}, see Appendix~\ref{AppendixNumerics} for more details.

\subsection{Zero magnetization \label{zeroM}}
We start by analyzing the spin-spin correlations along the $x$-direction.
We first consider the one-dimensional limit, $J_y = 0$, where the chains are uncoupled and form a Luttinger liquid.
Hence, the spin-spin correlations in the $z$-basis decay as \cite{Giamarchi2004}
\begin{equation}
	\left \langle \Sh^{z}_{j} \Sh^{z}_{j+x} \right \rangle = m^2 - \frac{K}{2 \pi^2} \frac{1}{x^2} + C_1 \cos(\pi [1+2m]x) \frac{1}{x^{2K}}.
	\label{eqLLCorrelationsZmag}
\end{equation}
Here, $m$ is the magnetization in the $z$-basis, $K$ is the Luttinger liquid parameter, and $C_1$ is a model dependent constant.
The value of $K$ corresponds to the strength of the spin-spin correlations, where higher values indicate faster decay of the correlations.

We fit Eq.~\eqref{eqLLCorrelationsZmag} to our DMRG data, first at $J_y = 0$, in one of the chains in the $x$-direction and extract an estimate $\tilde{K}$ for the LL parameter $K$; see Fig.~\ref{FigSix_correlations}(a) on the $x$-axis.
The value of $K$ decreases with decreasing value of $J_x^{\perp} / J_{x}^{z}$ because the AFM correlations become stronger with increasing value of the Ising term.
At the considered value, $J_y = 0$, corresponding to the 1D limit, where one can apply the Bethe ansatz, we obtain good agreement with the expected values from the LL-theory \cite{Giamarchi2004}.
Our results approach the exact value $K = 1/2$ at the isotropic AFM Heisenberg point $J_{x}^{\perp} / J_x^{z} = 1$ \cite{Giamarchi2004}.
For $J_{x}^{\perp} > J_x^{z}$ the LL parameter increases $K > 1/2$ as expected from the LL theory.
For $J_{x}^{\perp} < J_x^{z}$, the spins form a long-range AFM order; see also Sec.~\ref{PhysMech}.
This is reflected in $K \rightarrow 0$ as the system approaches the N\'eel state.

For low inter-chain coupling, $J_y \ll J^\perp_x, J^{z}_{x}$, the mixD system is close to an effective 1D description.
Hence, the correlations still follow the 1D form in Eq.~\eqref{eqLLCorrelationsZmag} on the length-scales considered in our numerics, because the effective interactions are renormalized and the 1D function starts to deviate only at longer distances.
We can thus use the 1D expression even for $J_y \neq 0$ since it yields the most accurate prediction of the transition point in the system.
We extract a LL-like parameter $\tilde{K}$, by using Eq.~\eqref{eqLLCorrelationsZmag}, and plot the results for different values of $J_y$ and $J_{x}^{\perp}$ in Fig.~\ref{FigSix_correlations}(a).
We provide exemplary fits in Appendix~\ref{AppendixNumerics}.
Our results show an extended region of long-range AFM order in the $x$-direction, stabilized by the inter-chain Ising interaction $J_y$ and destabilized by the $J^{\perp}_x$ spin-flip term.
To be more precise, we define the region where $ \tilde{K} \approx 1/2$ to be the boundary between the gapless disordered regime and the gapped ordered AFM regime, following the LL theory for the 1D limit.
Hence, the area where the extracted value drops below $\tilde{K} < 1/2$ corresponds to the AFM ordered regime.
There the correlator
\begin{equation}
    C^{z}_{AFM_{x}}(x) = (-1)^x
    \left \langle \Sh^{z}_{x_0, y_0} \Sh^{z}_{x_0+x, y_0}
    \right \rangle
    \label{eqSzSzStagCorrelator}
\end{equation}
is expected to saturate to a finite value in the very-long distance limit $\lim_{x \rightarrow \infty} (-1)^x \langle \Sh^{z}_{x_0, y_0} \Sh^{z}_{x_0+x, y_0} \rangle \neq 0$, see Appendix~\ref{AppendixNumerics}.
Since the inter-chain coupling $J_y$ aligns the spins along the $y$-direction, this region corresponds to AFM stripes sketched in Fig.~\ref{FigOne}(d).

As can be seen in Fig.~\ref{FigSix_correlations}(a) this area extends with $J_y > 0$, meaning that increasing inter-chain interactions further stabilizes AFM stripes for larger $J_{x}^{\perp}$.
This agrees with our experimental results in Fig.~\ref{FigTwo_OneHalfFilling}. 
In the \Zt LGT picture this results in an extended confined regime, where any \Zt charge excitation is confined.
For $J_{x}^{\perp} \gg J_y, J_x^{z}$, AFM stripes are destroyed and the system is disordered.
There, spinon excitations correspond to deconfined \Zt charges since the lack of AFM order does not induce a finite confining field $h$.

\subsection{Non-zero magnetization}
Next we study the system at finite magnetization, $m\neq 0$.
We start by considering the spin-spin correlations along the $x$-direction and extract the LL parameter $\tilde{K}$, again using Eq.~\eqref{eqLLCorrelationsZmag}.
The results presented in Fig.~\ref{FigSix_correlations}(b) show that $K$ increases with increasing $J_x^{\perp}$.
This is consistent with the LL theory in the limit when $J_y = 0$, because a lower value of $K$ generally indicates stronger interaction $J_x^{z}$ \cite{Giamarchi2004}.

Following the same arguments as in Sec.~\ref{zeroM}, we also analyze the LL-like parameter in the coupled regime $J_y > 0$.
The value of $\tilde{K}$ decreases even further with increasing coupling $J_y$ and decreasing $J_{x}^{\perp}$.
We note that for the Luttinger liquid $J_y = 0$, the value approaches $K \rightarrow 1/ 4$ for $J_{x}^{\perp} \rightarrow 0$, which is also obtained in the limit when the magnetization approaches zero $m \rightarrow 0$ and $J_x^{z} > J_x^{\perp}$ in 1D systems \cite{Giamarchi2004}.
These results indicate that inter-chain coupling significantly increases the interactions in the chain.
This is consistent with our experimental results in Fig.~\ref{FigThree_OneThirdFilling}, where the super-stripes melted by increasing the tunneling rate $t = J_\perp^{x} / 2$.

We analyze the spin-spin correlations further and consider the absolute value of long-range correlations, $ | \langle \Sh^{z}_{x_0, y_0} \Sh^{z}_{x_0+x, y_0}
     \rangle |$.
Now we fit the absolute value of our data with
\begin{equation}
    f(x) = |\cos(\pi \varphi x + \theta)  \le A + \frac{B}{x^{\alpha}} \r + C| ,
    \label{eqStripeFct}
\end{equation}
where $A, B, C, \theta$ and $\alpha$, are the fitting parameters.
A non-zero value of parameter $A$ indicates super-stripes, i.e., long-range incommensurate magnetic order.
In addition, $\varphi$ determines the period of the stripes.
Other parameters are used for technical purposes to perform the fit.
Here we analyze $A$, for which we show the results in Fig.~\ref{FigSix_correlations}(c); see also Appendix~\ref{AppendixNumerics} for more details.
We obtain a finite value of $A$ for $J_y >0$ and $J_{x}^{\perp} \ll J_{x}^{z}$, consistent with the decrease of the LL-like parameter $\tilde{K}$ in Fig.~\ref{FigSix_correlations}(b).
These results signal strong spin correlations along the $x$-direction that provide strong evidence for incommensurate magnetism and super-stripes for $J_y > 0$; see Fig.~\ref{FigOne}(e).

\begin{figure}[t]
\centering
\epsfig{file=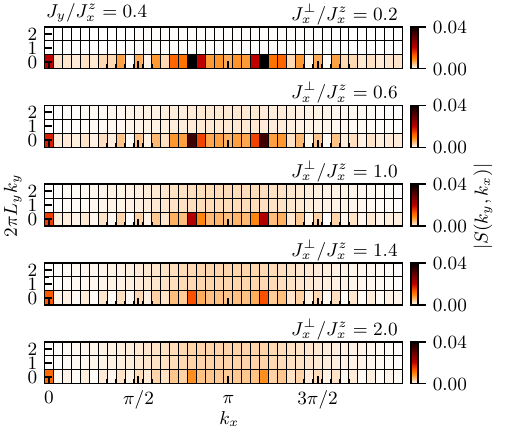}
\caption{
Numerical results for the Fourier transformation of the spin-spin correlations Eq.~\eqref{eqFourier2DSpinSpin} for finite magnetization $m = 0.1$. Sharp peaks of $S(k_x, k_y)$ at momenta $ k_x = \pi (1 \pm 2m)$, $k_y = 0$, signal the formation of super-stripes. The periodicity of the stripes is proportional to the total magnetization of the chains $m$. Stripe order becomes weaker with increasing $J^{\perp}_x$ term (top to bottom). The peaks become less pronounced and $S(k_x, k_y)$ acquires non-zero contributions also in the $k_y$-direction, corresponding to reduced inter-chain correlations.
The system size is $L_x = 40$ and $L_y = 3$.
}
\label{FigSeven_FourierGS}
\end{figure}

So far we studied spin-spin correlations only along the $x$-direction.
To study the super-stripe phase explicitly for $m \neq 0$, we analyze spin-spin correlations in the full 2D system.
We consider the spin-spin correlator 
\begin{equation}
    C^z(x , y) =  \left \langle \Sh^{z}_{x_0, y_0} \Sh^{z}_{x, y} \right \rangle,
    \label{eq2DSpinSpin}
\end{equation}
where we select a reference spin in the middle of the system $x_0 = 20$ and $y_0 =2$, and calculate the correlations with other sites.
We note that this correlation function is closely related to the experimental correlator Eq.~\eqref{eq:d_correlation}, up to an offset due to the mapping between hard-core bosons and spins.
Next, we perform a Fourier transformation of the correlator Eq.~\eqref{eq2DSpinSpin}, defined as
\begin{multline}
    S(k_x,k_y) = \\
    \frac{1}{L_x L_y} \sum_{x = 0}^{Lx-1} \sum_{y=0}^{Ly-1}
    e^{-i (x - x_0) k_x} e^{-i (y - y_0) k_y} C^z(x , y) .
    \label{eqFourier2DSpinSpin}
\end{multline}
Here we discretized the Fourier modes as $\Delta k_x = \frac{2 \pi}{L_x}$ and $\Delta k_y = \frac{2 \pi}{L_y}$.

Super-stripes break the translational symmetry because spins form a periodic standing wave pattern, see Fig.~\ref{FigOne}(e).
This can be probed in the spin-spin Fourier transformation Eq.~\eqref{eqFourier2DSpinSpin}, which contains peaks related to the super-stripe frequency.
We observe such peaks at $k_x = \pi \le 1 \pm 2m \r$, $k_y = 0$ when $J^{\perp}_{x} / J^{z}_{x} \lesssim 0.5$; see Fig.~\ref{FigSeven_FourierGS}.
The peaks decrease and the super-stripe order is melted for $J^{\perp}_{x} / J^{z}_{x} \gtrsim 0.5 $, although in our finite-size system some features remain visible at the $k$-values where super-stripes peak.
In addition, Eq.~\eqref{eqFourier2DSpinSpin} acquires nonzero contributions also in the $y$-direction since the correlations between individual chains decrease.

The super-stripe regime corresponds to the confined regime in the \Zt basis because aligned AFM domain walls (spinons) in the $y$-direction correspond to aligned \Zt charges, which form line-like objects in the $y$-direction; see Fig.~\ref{FigOne}(e).
With increasing $J^{\perp}_{x}$ such order is destroyed and \Zt charges (i.e. spinons) deconfine, leading to the absence of magnetic order.

Combining the 1D spin-spin correlation result in Fig.~\ref{FigSix_correlations} and the Fourier transform of the 2D correlations in Fig.~\ref{FigSeven_FourierGS} allows us to qualitatively sketch the phase diagram at finite magnetization in Fig.~\ref{FigOne}(e). 
We observe the super-stripes at finite magnetization, stabilized by the inter-chain coupling $J_{y}$, consistent with discussion in Sec.~\ref{PhysMech}.
Furthermore, our numerical results are qualitatively consistent with our experimental results, where the system possesses also long-range interactions, which we truncate in the numerical calculations.

\section{Meson gas at finite temperature \label{FiniteTnumerics}}
Next we consider the mixD XXZ model Eq.~\eqref{eqMixedDXXZModel} at finite temperature.
In order to study the confined phase at finite temperature we fully embrace the \Zt LGT picture Eq.~\eqref{eq_mappedZ2LGT} to which we map our mixD system in Sec.~\ref{MappingToLGT}.
By using numerical calculations we show that the systems features commensurate AFM magnetism at finite temperature and low magnetization, which corresponds to the meson gas in the \Zt LGT picture \cite{Grusdt2020}.

In order to obtain finite temperature results, we employ the quantum purification scheme where we attach an ancilla lattice site to every physical site \cite{Zwolak2004, Feiguin2005, Feiguin2013, Nocera2016}.
We then implement the maximally entangled state between ancilla and physical sites, which corresponds to the infinite temperature state $\beta = 1 / T = 0$.
Hereafter we perform imaginary time evolution \cite{Paeckel2019} to obtain the finite temperature results.
In addition, we add a field in the $z$-direction to tune the magnetization; see Appendix~\ref{AppendixNumerics} for more details on the numerical calculations.
Finite temperature calculations are performed by employing  \textsc{SyTen} \cite{hubig:_syten_toolk, hubig17:_symmet_protec_tensor_network}.

\begin{figure}[t]
\centering
\epsfig{file=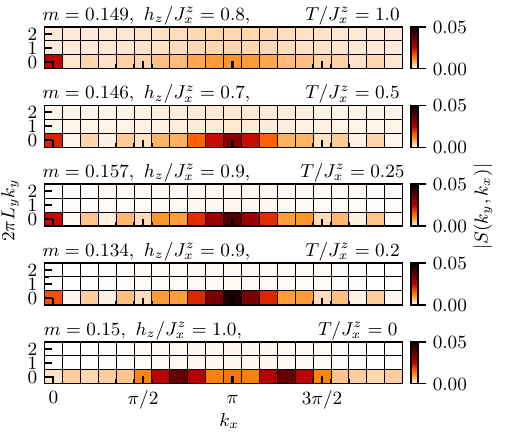}
\caption{
Numerical results for the Fourier transformation Eq.~\eqref{eqFourier2DSpinSpin} of the spin-spin correlations Eq.~\eqref{eq2DSpinSpin} at approximately $|m| \approx 0.15$ magnetization, at different temperatures $T$.
Two peaks observed in the ground state $T = 0$ (bottom), which signal the super-stripe regime, merge into a single peak at low temperature values $ T \lesssim 0.5 J_x^{z}$, which signals the meson gas regime.
At higher temperature $ T \gtrsim 0.5 J_x^{z}$ the peak broadens and the meson gas is diluted into a simple chargon gas.
The system size is $L_x = 20$ and $L_y = 3$ and the parameters are set to $J^{\perp}_{x} / J^{z}_{x} = 0.4$ and $J_{y} / J^{z}_{x} = 0.8$ in our MPS simulations.
}
\label{FigEight_mesonG}
\end{figure}

\subsection{Melting the super-stripe order}
Super-stripe order in the mixD XXZ model maps to the confined phase in the \Zt LGT picture, where the \Zt charges form line-like objects along the $y$-direction; see Fig.~\ref{FigOne}(e).
In the ground state, $T = 0$, the \Zt charges can be considered to be locked into stripes.

At finite temperature $T > T_s$, the super-stripe order is destroyed.
However, the system exhibits a confining \Zt field on the mean-field level as long as the staggered magnetization remains non-zero in the spin picture.
This occurs provided the interactions among spins are sufficiently strong to induce a commensurate long-range AFM order for low temperature $0 < T_s < T < T_{\rm AFM}$.
This mediates confinement on the mean-field level between the chains because the spontaneous magnetic order induces a finite value of the \Zt field $h \neq 0$, discussed in Sec.~\ref{PhysMech}.
By melting the long-range super-stripe order we thus observe confinement of spinon pairs, i.e., a gas of confined mesons: these correspond to the magnon (spin-wave) excitations of the commensurate AFM.
The confined state is ultimately destroyed when the temperature becomes larger than the energy scale of the Ising interactions, $T > T_{\rm AFM}$.
There, any long-range AFM order is destroyed and spinons are deconfined, resulting in the deconfined chargon gas.

To probe the temperature-induced transition from super-stripe to meson gas, we consider the Fourier transformation of the spin-spin correlation, Eq.~\eqref{eqFourier2DSpinSpin}, and examine whether the super-stripe peaks at $k_x = \pi (1 \pm 2m )$
are replaced by a single pronounced peak at $k_x = \pi$.
Such behavior indicates that the stripe order is destroyed and that a spontaneous AFM order developed.
In the regime where magnetization is finite, this results in a gas of confined mesons.
We thus consider the system at finite magnetization $m \neq 0$, which also ensures a sufficient number of spinons.

The results for $m = 0.15$ for different temperatures $T$ can be seen in Fig.~\ref{FigEight_mesonG}.
In the ground state, $T = 0$, two peaks at $k_x = \pi \le 1 \pm 2m \r $ indicate the super-stripe regime.
At low temperature $0 < T \lesssim 0.5 J_x^z$, we observe a single peak at $k_x = \pi$.
Since the magnetization is finite, $m = 0.15$, we identify this feature as the key signature of the meson gas, where partons confine into mesons in an AFM background.
At higher temperatures $T \gtrsim 0.5 J_x^z$, the peak at $k_x = \pi$ broadens, indicating weaker AFM correlations resulting in the decreased value of the \Zt confining field $h$.
Hence, the system eventually transitions to a free chargon gas (deconfined spinons) since the AFM correlations become weaker, consistent with our analytical predictions discussed in Sec.~\ref{PhysMech}.

These features are also observed at higher magnetization, where the peak becomes broader at higher magnetization already at low temperature; see Appendix~\ref{MoreFiniteTspinCorrelations}.
Based on these observations, we qualitatively sketch the phase diagram presented in Fig.~\ref{FigOne}(f).
We expect that the meson gas can be formed only at low magnetization, i.e., that the super-stripes are melted directly into the chargon gas for higher magnetization $m$, where spontaneous AFM order cannot be realized.
We note that similar physics is also observed in a mixed-dimensional $t-J_{z}$ model studied in Ref.~\cite{Grusdt2020}, where the authors employ large-scale quantum Monte Carlo simulations and parton mean-field theory.
There, holes in an AFM ordered spin configuration act as \Zt charges, which form stripes at low temperature and confine into mesons at low doping above the stripe melting temperature.
Furthermore, MPS simulations do not suffer from the sign problem. We can thus investigate very low temperatures and the ground state $T=0$.

\subsection{String-length distributions} \label{SectionV_strings}
To further study the confined meson gas, we consider string and anti-string length distributions as defined in the \Zt LGT picture.
In the confined regime strings are on average much shorter than anti-strings, see Fig.~\ref{FigOne}.
Hence, string and anti-string length histograms feature a bimodal distribution in the confined regime \cite{Kebric2023FiniteT}.
In the super-stripe regime or the deconfined chargon gas, the distributions of string and anti-strings are the same.

\begin{figure}[t]
\centering
\epsfig{file=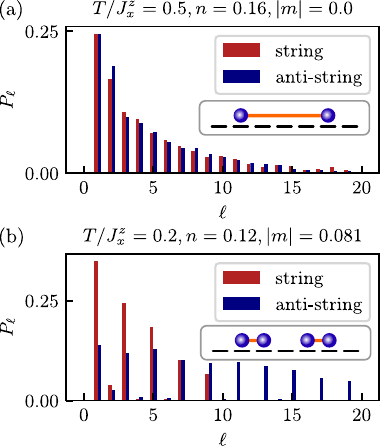}
\caption{
Numerical results for the string and anti-string length distributions from MPS snapshots.
(a) At high temperature $T / J_{x}^{z} = 0.5$ the string and anti-string length distributions are similar to each other which indicates a deconfined phase.
(b) At lower temperature $T / J_{x}^{z} = 0.2$ the string-length has a higher peak at $\ell = 1$ than the anti-string length distribution. In addition the anti-string length distribution has a longer tail. This shows that the mesons are confined at lower temperature and form a meson gas.
The length of the cylinder is $L_x = 20$ and the circumference is $L_y = 3$.
Furthermore, we set the parameters to $J^{\perp}_{x} / J^{z}_{x} = 0.4$ and $J_{y} / J^{z}_{x} = 0.8$. With $n$ we denote the average number of partons in the \Zt LGT picture.
}
\label{FigNine_strings}
\end{figure}

To study the string and anti-string length distributions numerically, we sample snapshots from our MPS \cite{Buser2022, Ferris2012}.
We then examine each snapshot and record the lengths between even-odd and odd-even particles, which reflect the length of the strings and anti-strings in the \Zt LGT picture \cite{Kebric2023FiniteT}.
The string and anti-string length distributions for low magnetization $m$, are presented in Fig.~\ref{FigNine_strings}.
We observe that both distributions look similar at high temperature $T / J_{x}^{z} = 0.5$, suggesting that there is no confinement of partons into mesons and we simply have deconfined partons.
This corresponds to the deconfined chargon gas.
However, at lower temperature $T / J_{x}^{z} = 0.2$, the string and anti-string distributions differ.
The strings have a more pronounced peak at $\ell = 1$ than the anti-strings and the anti-string distribution has a long tail.
This suggests that partons confine into mesons and the system forms a meson gas.

When analyzing snapshots, we generally assume that our system starts and ends with an anti-string $\tau^{x}_{\langle 0, 1 \rangle}, \tau^{x}_{\langle L, L+1 \rangle} = +1$, and therefore the first domain wall in the chain is the start of the first string.
Since the effective confining field arises on a mean-field level, the orientation of the \Zt electric field value is not exactly determined.
As a result, the identity of strings and anti-strings might be inverted.
We mitigate this problem by computing the value of the staggered magnetization $m_s$.
In the confined state, the strings are on average shorter than the anti-string lengths, which results in a nonzero value of the staggered magnetization.
Based on the sign of $m_s$ we can thus determine whether our assumption that the system started with an anti-string was correct.
In the case where the total staggered magnetization is positive, we exchange the identity of strings and anti-strings for that particular snapshot.
Such classification of strings and anti-strings works best deep within the different phases, as the fluctuations near the transitions can introduce additional noise.
We use the same procedure when analyzing the experimental snapshots in Fig.~\ref{FigFour_ExperimentStrings}(a) and Fig.~\ref{FigFour_ExperimentStrings}(c).

We note that the effective \Zt LGT Eq.~\eqref{eq_mappedZ2LGT}, does not have a global $U(1)$ parton number conservation due to pair creation and annihilation terms.
This results in parton number fluctuation and contributes to longer tails in both distributions \cite{Kebric2026MF}.
The overall distribution also depends on the effective density of partons in the system $n$.
This cannot be directly fixed in our numerical setup and depends on the temperature $T$, magnetization $m$, and the parameter values of the system.
For higher fillings the distributions are narrower and it is generally harder to distinguish between the confined and deconfined state \cite{Kebric2026MF, Kebric2023FiniteT}.

\subsection{Parton spacing distributions} \label{SecV_partonSpacing}
\begin{figure}[t]
\centering
\epsfig{file=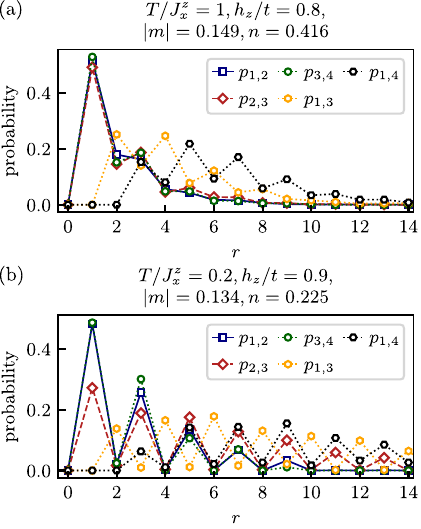}
\caption{Probability distributions of the distance between partons $p_{a, b}(r)$, obtained from the numerical MPS calculations for finite magnetization $|m| \approx 0.15$.
(a) For high temperature $T = J_{x}^{z}$, NN parton distance distributions are qualitative the same $p_{1,2}(r) \approx p_{2,3}(r) \approx p_{3,4}(r)$, with a peak at $r = 1$, as expected for the deconfined chargon gas phase.
(b) For low temperature $T / J_{x}^{z} = 0.2$, distributions between odd-even and even-odd partons differ significantly $p_{1,2}(r) \approx p_{3,4}(r) \neq p_{3,4}(r)$, as partons confine into mesons, where strings are on average shorter than the anti-strings.
The parameters in this plot are $J^{\perp}_{x} / J^{z}_{x} = 0.4$, $J_{y} / J^{z}_{x} = 0.8$, and the cylinder has a length of $L_x = 20$ and circumference $L_y = 3$.
}
\label{FigTen_MPSpartonDist}
\end{figure}

Finally, we study the parton spacing distributions $p_{a,b}(r)$, in our numerical simulations.
We consider high and low temperature regimes at a finite magnetization $|m| \approx 0.15$, which corresponds to the chargon and meson gas regimes, respectively.
For high temperature ${T / J_x = 1}$, the nearest-neighbor distributions have qualitatively same shapes, $p_{1,2}(r) \approx p_{2,3}(r) \approx p_{3,4}(r)$, with peaks at $r = 1$, see Fig.~\ref{FigTen_MPSpartonDist}(a).
This is expected for the free chargon gas regime, where the \Zt charges are deconfined.
For low temperature $T / J_x = 0.2$, meson gas exhibits different behavior where $p_{1,2}(r=1) \approx {p_{3,4}(r=1)} > p_{2,3}(r=1)$, as strings become on average much shorter than the anti-strings, see Fig.~\ref{FigTen_MPSpartonDist}(b).
This is the key microscopic signature of the meson gas.
We note that we determine the identity of strings and anti-strings based on the total staggered magnetization in the same way as for the string and anti-string length histograms in Fig.~\ref{FigNine_strings}.
Hence, we relabel every parton as $s \rightarrow s+1$ in case the staggered magnetization of the whole system is positive; see Appendix~\ref{AppendixPartonSpacing} for details.

\section{Discussion and Conclusion \label{Conclusion}}
In this work we studied the confined and deconfined phases in an array of coupled 1D \Zt LGTs.
We experimentally simulate a Bose-Hubbard model using erbium atoms in a quantum gas microscope, which feature tunable dipolar interactions.
We consider the hard-core limit, with particle hopping limited to a single spatial direction that can be mapped to the coupled arrays of \Zt LGTs.
We observed stable stripe and super-stripe phases, formed by line-like objects created by particles aligned along one of the spatial directions that are periodically spaced in the other direction.
We are able to melt the stripes by increasing the hopping amplitude in our system, which corresponds to a confinement-deconfinement transition in the \Zt LGT picture.

In order to provide a theoretical description of our results, we map the Bose-Hubbard Hamiltonian to a spin-$1/2$ system, which yields arrays of XXZ chains, coupled by an Ising interaction.
This mixD XXZ model is obtained as an approximation, where beyond nearest neighbor dipolar interactions are truncated.
Spin domain walls (spinons) in the mixD model map to \Zt charges on a dual lattice.
In the ground state, inter-chain coupling aligns the \Zt fields of neighboring \Zt LGT arrays to minimize their energy.
\Zt charges thus form striped phases, with the partons locked to their corresponding stripes.

Furthermore, a confining electric field term emerges in our system, provided that the coupled XXZ chains form a spontaneous long-range AFM order.
This occurs at a finite temperature after the stripe phase is destroyed by thermal fluctuations, as long as the magnetization and the temperature remain low enough.
Finite electric field term induces a linear confining potential for the \Zt strings that connect the partons, which form a gas of confined mesons (bound parton pairs).
Hence, the disordered regime in the spin picture corresponds to the deconfined phase in the \Zt LGT picture.

We use MPS based numerical techniques to simulate the mixD XXZ model and construct a comprehensive phase diagram of the system.
We confirm that the system in the ground state forms a gapped AFM stripe state at zero magnetization, and a super-stripe state at finite magnetization, which is identified by peaks in the Fourier transformation of the spin-spin correlations at the incommensurate magnetization values.
Striped phases are destroyed by increasing the XX term of the mixD model, which is in qualitative agreement with our experimental results, where we simulate a full Bose-Hubbard Hamiltonian with dipolar interactions.

Our numerical calculations at finite temperature and magnetization show the existence of a meson gas, where spin-spin correlations feature a single peak at $k_x = \pi$, indicating spontaneous AFM order.
By increasing the temperature, the peak becomes less pronounced since the meson gas melts into a deconfined phase, i.e., a chargon gas of free partons.
We also sample snapshots from our MPS and plot the string and anti-string length distributions, as well as the parton spacing distributions.
We find that the system has confining features at low temperature, where we find the meson gas.
In our experiment, the meson gas features are less pronounced, which we attribute to long-range dipolar interactions.
Long-range interactions map to parton-parton interactions in the \Zt LGT picture that tend to stabilize the super-stripe structures.

Our results show that transition between the super-stripe and disordered state can be understood in terms of a confinement-deconfinement transition in the \Zt LGT picture.
We show that mixed-dimensional systems can be used as a platform to study confined phases in \Zt LGTs with matter.
We demonstrate that striped phases are stable in experimental settings even when long-range dipolar interactions are included.
While we report first experimental signatures of meson formation, achieving a stable confined meson gas in an experimental setting remains a challenge for future studies, for which our first experimental exploration in this work provides an important stepping stone.

\section*{Acknowledgements}
We thank Lukas Homeier, Adam Kaufman, Mattia Moroder, Felix Palm, Lode Pollet, Ana Maria Rey, and Henning Schlömer for fruitful discussions.
The lab is grateful for the early contributions to building the experiment from A. Krahn, A. Hebert, R. Groth, G. Phelps, S. F. Ozturk, S. Ebadi, S. Dickerson, and F. Ferlaino.
This project has received funding from the European Research Council (ERC) under the European Union’s Horizon 2020 research and innovation programm (Grant Agreement no 948141) — ERC Starting Grant SimUcQuam, and by the Deutsche Forschungsgemeinschaft (DFG, German Research Foundation) under Germany's Excellence Strategy -- EXC-2111 -- 390814868. 
The experimental work is supported by U.S. Department of Energy Quantum Systems Accelerator DE-AC02-05CH11231, National Science Foundation Center for Ultracold Atoms PHY-1734011, Army Research Office Defense University Research Instrumentation Program W911NF2010104, Office of Naval Research Vannevar Bush Faculty Fellowship N00014-18-1-2863, Gordon and Betty Moore Foundation Grant GBMF11521, and Defense Advanced Research Projects Agency Optimization with Noisy Intermediate-Scale Quantum devices W911NF-20-1-0021. A.D. acknowledges support from the NSF Graduate Research Fellowship Program (grant DGE2140743).

\section*{Competing interests}
M.G. is a cofounder and shareholder of QuEra Computing. All other authors declare no competing interests.

\appendix

\begin{figure*}[t]
\centering
\epsfig{file=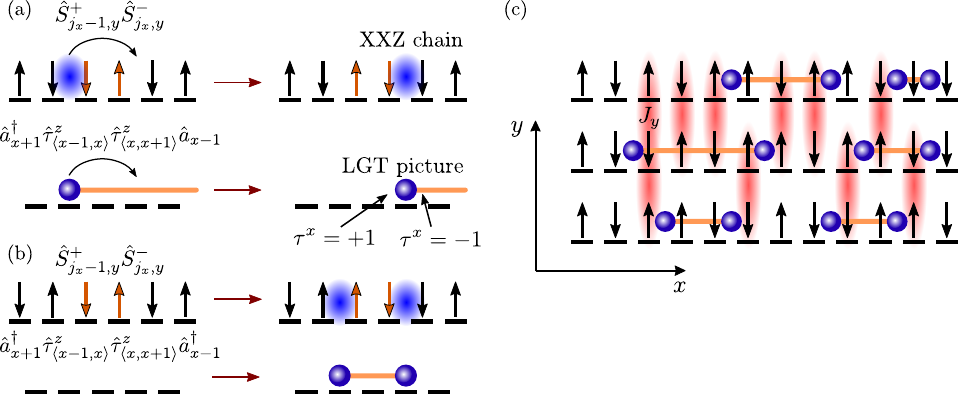}
\caption{
Mapping of the mixed-dimensional XXZ chains to a a \Zt lattice gauge theory, where we define hard-core bosons on the links between two aligned spins.
(a) Spin exchange term induces hopping over two lattice in a configuration where a domain wall already exists.
(b) In a spin configuration without any domain walls a spin flip operator creates two domain walls, which in the \Zt LGT basis results in a parton pair over two lattice sites. Pair annihilation can also occur in the opposite configuration.
(c) The inter-chain Ising interaction, which drives the transition to the super-stripe regime in the ground state, induces confinement of domain wall pairs on the mean field level.
}
\label{FigAppLGTMapping}
\end{figure*}

\section{Mapping of the mixed-dimensional XXZ chain to a lattice gauge theory}\label{AppendixLGTmapp}

Here we provide additional details on mapping the mixD XXZ model to an array of coupled one-dimensional \Zt LGTs.
We start by briefly reiterating the definitions explained in the main text.
Hard-core bosons (partons) are defined on dual lattice sites 
\begin{equation}
    \hat{n}_{x, y} = \ad_{x, y} \a_{x, y} =
    \frac{1}{2} \le 1 + 4 \Sh^{z}_{j_x-1, y}\Sh^{z}_{j_x, y} \r .
    \label{eqSpinonsAsHCbosons}
\end{equation}
Here, $( j_x, y )$ corresponds to the original mixD lattice site along the $x$ and $y$ direction, and $(x, y)$ corresponds to the dual lattice sites; see Fig.~1(c) in the main text.
Note that labeling along the $y$ directions remains the same.
The above definition comes from restricting the Hilbert space to the physical sector via the Gauss law, Eq.~\eqref{eqGaussLawLGT}.
In addition, we define the \Zt electric field in terms of the spins as
\begin{equation}
    \tauX_{\langle x, x+1 \rangle , y} \equiv \tauX_{\vecj} =  2 (-1)^{j_x} \Sh_{\vecj}^{z} .
    \label{eqAppTauXDef}
\end{equation}
Using Eq.~\eqref{eqSpinonsAsHCbosons} and Eq.~\eqref{eqAppTauXDef} allows us to map individual terms in the mixD XXZ model to the \Zt LGT.

The spin flip term proportional to $J_{x}^{\perp}$ maps to two different terms in the \Zt LGT basis, depending on the spin configuration.
It can move a domain wall by two lattice sites, see Fig.~\ref{FigAppLGTMapping}(a).
In the second case it can create or annihilate a pair of domain walls as depicted in Fig.~\ref{FigAppLGTMapping}(b).
We also note that the spins have to be anti-aligned for this term to be non-zero, which means that in the \Zt LGT picture, the dual site $ (x, y) $ has to be empty.
We can thus map the spin flip term as
\begin{widetext}
\begin{multline}    
    \le \Sh^{+}_{j_x, y} \Sh^{-}_{j_x-1, y} + \Sh^{+}_{j_x-1, y} \Sh^{-}_{j_x, y} \r
    \rightarrow \\
    \le \ad_{x-1, y } \tauZ_{\langle x-1, x \rangle , y} \tauZ_{\langle x, x+1 \rangle , y} \a_{ x+1, y}
    + \ad_{ x-1, y } \tauZ_{\langle x-1, x \rangle , y} \tauZ_{\langle x, x+1 \rangle , y} \ad_{x+1, y} + \hc \r 
    \le 1 -  \ad_{x, y } \a_{x,y }  \r .
    \label{eqHoppingMappingToLGT}
\end{multline}
\end{widetext}

The Ising term along the $x$-direction maps directly to a chemical potential term as
\begin{equation}
    \Sh^{z}_{j_x-1, y}\Sh^{z}_{j_x, y} \rightarrow
    \frac{1}{4} \le 2 \hat{n}_{x, y} -1 \r ,
    \label{eqChemicalMappingToLGT}
\end{equation}
which can be used to control the filling in the chains.

The Ising interaction in the $y$-direction is rewritten in terms of the \Zt electric field by taking into account the definition in Eq.~\eqref{eqAppTauXDef}
\begin{multline}
     \Sh^{z}_{j_x, y}\Sh^{z}_{j_x, y+1} \rightarrow \\
     \frac{1}{2} (-1)^{j_x} \tauX_{\langle x, x+1 \rangle , y} 
     \frac{1}{2} (-1)^{j_x} \tauX_{\langle x, x+1 \rangle , {y+1}} \\
     = \frac{1}{4}  \tauX_{\langle x, x+1 \rangle , y} \tauX_{\langle x, x+1 \rangle , {y+1}} .
    \label{eqChainCouplingMappingToLGT}
\end{multline}

By combining all of the terms we obtained coupled arrays of one-dimensional \Zt lattice gauge theory in the main text
\begin{widetext}
\begin{multline}
    \H_{LGT} = \frac{J_{x}^{\perp}}{2} \sum_{x, y} \le \ad_{x-1, y} \tauZ_{\langle x-1,x \rangle , y} \tauZ_{\langle x, x+1 \rangle , y} \a_{ x+1, y } + \hc \r
    \le 1 -  \hat{n}_{x, y}  \r \\
    + \frac{J_{x}^{\perp}}{2} \sum_{x, y} \le \ad_{ x-1, y} \tauZ_{\langle x-1, x \rangle , y} \tauZ_{\langle x, x+1 \rangle , y} \ad_{ x+1, y} + \hc \r
    \le 1 -  \hat{n}_{x, y} \r \\
    + \frac{J_{x}^{z}}{4} \sum_{x, y} \le 2 \hat{n}_{x, y} - 1 \r 
    - \frac{J_y}{4} \sum_{x, y} \tauX_{\langle x, x+1 \rangle , y} \tauX_{\langle x, x+1 \rangle , {y+1}} + ...
    \label{eqMappedLGT}
\end{multline}
\end{widetext}

Some comments are in order. 
First of all, particles hop over two lattice sites, with the condition that the site in between has to be empty.
Similarly, pair-creation and annihilation terms also act across two lattice sites.
The pair creation/annihilation term also explicitly breaks the U(1) parton number conservation \cite{Kebric2026MF}.
Despite hopping over two lattice sites, this \Zt LGT exhibits physics similar to the standard one-dimensional \Zt LGT \cite{Borla2020PRL, Kebric2021, Kebric2023FiniteT}.

The main difference from the standard one-dimensional \Zt LGT is the last term that induces confinement and arises from the inter-chain Ising coupling $J_y$.
To understand the confinement in Hamiltonian Eq.~\eqref{eqMappedLGT}, we simplify the analysis by considering the \Zt electric field in neighboring chains on a mean-field level.
In the deconfined regime we can assume that strings and anti-strings are equivalent and their average lengths are the same.
In the deconfined regime the average expectation value of the \Zt electric field terms is therefore zero, $\langle \tauX_{\langle x, x+1 \rangle , y} \rangle = 0$.
Contrarily to the deconfined case, the average expectation value of the electric field term would be non zero in the confined case $\langle \tauX_{\langle x, x+1 \rangle , y} \rangle > 0$, since particle pairs bind into mesons which are on average shorter than the distance between particles connected with anti-strings; see Fig.~\ref{FigAppLGTMapping}(c).
This term can thus be rewritten on the mean field level as
\begin{equation}
    \frac{J_y}{4} \sum_{x, y} \tauX_{\langle x, x+1 \rangle , y} \tauX_{\langle x, x+1 \rangle , y+1}
    \rightarrow - h \sum_{x,y} \tauX_{\langle x, x+1 \rangle , y},
\end{equation}
where we defined the electric field value as 
\begin{equation}
    h = \frac{J_y}{8} \le \big \langle \tauX_{\langle x, x+1 \rangle , y+1} \big  \rangle + \big \langle \tauX_{\langle x, x+1 \rangle , y-1} \big  \rangle \r.
    \label{eqConfhMF}
\end{equation}
The expectation value of the electric field term is in addition directly related to the staggered magnetization of the chain
\begin{equation}
    \langle \tauX_{\langle x, x+1 \rangle , y} \rangle = 2 m_s^{y} =  \frac{2}{L_x} \sum_{j_x} (-1)^{j_x} \langle \psi | \Sh^{z}_{\vecj} | \psi \rangle.
\end{equation}
Uncoupled chains $J_y = 0$ will always be deconfined.
Furthermore, systems where the staggered magnetization is zero will also be deconfined.

Finally, we comment on long-range interactions in the \Zt LGT basis.
Interactions along the $y$ direction map as
\begin{equation}
    V_{0, d \bm{e}_y} \Sh^{z}_{j_x, y} \Sh^{z}_{j_x, y+d} \rightarrow 
    V_{0, d \bm{e}_y} \tauX_{\langle x, x+1 \rangle , y} \tauX_{\langle x, x+1 \rangle , {y+d}},
\end{equation} where $d$ is the distance between spins.
Such interactions thus renormalize the electric field term on the mean-field level, Eq.~\eqref{eqConfhMF}.
Long-range interactions along the $x$-direction can be understood in terms of additional density-density interactions among the \Zt charges.
As an example we consider the next nearest-neighbor interaction $\propto V_{2 \bm{e}_x, 0} \Sh^{z}_{j_x, y} \Sh^{z}_{j_x + 2, y}$, which can be mapped to 
\begin{multline}
    \Sh^{z}_{j_x, y} \Sh^{z}_{j_x + 2, y} \rightarrow \\
    \rightarrow \frac{1}{2} (-1)^{j_x} \tauX_{\langle x, x+1 \rangle , y} \frac{1}{2} (-1)^{j_x + 2} \tauX_{\langle x+2, x+3 \rangle , y} \\
    = \frac{1}{4} \tauX_{\langle x, x+1 \rangle , y} \tauX_{\langle x+1, x+2 \rangle , y} \tauX_{\langle x+1, x+2 \rangle , y} \tauX_{\langle x+2, x+3 \rangle , y} \\
    = \frac{1}{4} (1-2\hat{n}_{x,y}) (1-2\hat{n}_{x+1,y}) ,
    \label{eqNNIsingInx}
\end{multline}
where we used relation Eq.~\eqref{eqGaussDenField}, between the \Zt electric field orientation and the \Zt charges.
Using the same approach we can show that a third-neighbor interaction $d = 3$ results in slightly more complicated density-density interactions
\begin{multline}
    \Sh^{z}_{j_x, y} \Sh^{z}_{j_x + 2, y} \rightarrow \\
    - \frac{1}{8} (1-2\hat{n}_{x,y}) (1-2\hat{n}_{x+1,y}) (1-2\hat{n}_{x+2,y}).
    \label{eq3NNint}
\end{multline}
Such interactions can stabilize ($d = 2$) or destabilize ($d = 3$) translational symmetry breaking states for commensurate parton fillings providing that they are strong enough.
For the case $ d = 2$ we obtain repulsive NN interaction, Eq.~\eqref{eqNNIsingInx}, and for $d = 3$ we obtain NN attractive interactions and a 3-body NN repulsive interaction, Eq.~\eqref{eq3NNint}.

\begin{figure}[t]
\centering
\epsfig{file=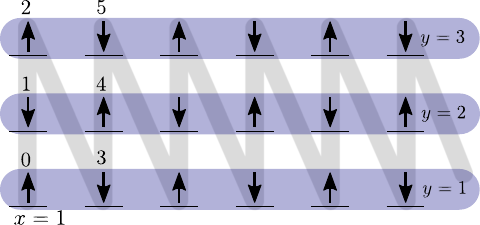}
\caption{
A sketch of a one-dimensional MPS chain winding through a two-dimensional lattice of size $L_y = 3$. As a result the nearest-neighbor interactions on the two-dimensional lattice correspond to $L_y$-th neighbour interaction in the one-dimensional MPS chain.
}
\label{FigAppMPSonCylinder}
\end{figure}

As an example we mention that the NN repulsion, Eq.~\eqref{eqNNIsingInx}, can stabilize a simple parton Mott state at half filling $n = 1/2$ when it is stronger than the hopping amplitude $V > 2t$ \cite{KebricNJP2023, Giamarchi2004}.
In addition, interplay between local NN charge repulsion and non-local confining fields can result in interesting phases of matter resembling the parton-plasma phase \cite{KebricNJP2023}.
However, the system remains confined on the long length-scale.
In our case these interactions are generally too weak to stabilize Mott states on their own.
In addition, we obtained repulsive as well as attractive interactions, which compete with each other.
Hence, long-range interactions can result in marginally altered stability of the translational symmetry breaking super-stripe phases.
The mechanism of confinement remains unaffected, and only the extent of the super-stripe phase region can be altered.
Finally, in the experimental setting, studying the effect of long-range interactions is further complicated by the presence of disorder and finite temperature.

\begin{figure*}[t]
\centering
\epsfig{file=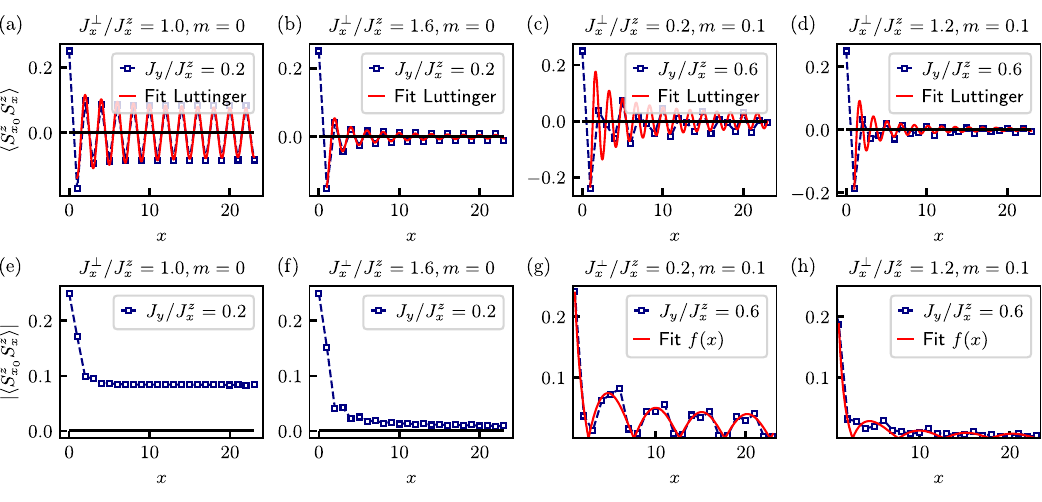}
\caption{
Numerical results (DMRG) of the spin-spin correlations along the $x$-direction in the mixD XXZ model, Eq~\eqref{eqMixedDXXZModel}.
(a-d) Fits of the spin-spin correlations with Eq.~\eqref{eqLLCorrelationsZmag}, for different values of $J_x^{xy}$, $J_y$, and magnetization $m$ from which we extract the LL-like parameter $\tilde{K}$.
(e-f) Absolute value of the spin-spin correlations for zero magnetization $m = 0$, which correspond to $C^{z}_{AFM_{x}}(x)$, Eq.~\eqref{eqSzSzStagCorrelator} in the main text, saturate to a finite value in the AFM stripe regime (e), and decay to zero in the disordered regime (f).
(g-h) Absolute value of the spin-spin correlations for finite magnetization $m = 0.1$, fit with Eq.~\eqref{eqStripeFct}.
}
\label{FigAppGS_corrFits}
\end{figure*}

\section{Numerical calculations}\label{AppendixNumerics}
\subsection{Ground state calculations}
We use DMRG \cite{Schollwoeck2011, White1992}, more precisely a MPS toolkit \textsc{SyTen} \cite{hubig:_syten_toolk, hubig17:_symmet_protec_tensor_network} to numerically calculate the ground state of the mixD XXZ chain.
We consider a one-dimensional MPS chain, which we wind on a cylinder as depicted in Fig.~\ref{FigAppMPSonCylinder} to emulate a two-dimensional spin lattice.
As a result, the nearest-neighbor interactions on the 2D square lattice result in $L_y$-th neighbor interaction in the actual MPS implementation.
This requires greater care when converging our results, and we typically allow for the bond dimension to grow up to $\chi = 2048$.

If not stated otherwise, we use a global U(1) symmetry of the system, where we can choose a desired magnetization sector $m$.
In addition, we use open boundary conditions in the $x$-direction and periodic boundary conditions in the $y$-direction.

\subsection{Spin-spin correlations along the \texorpdfstring{$x$}{x}-direction}
Here we provide more details on the spin-spin correlations along the $x$-direction.
We first show typical fits of the numerical data with the Luttinger liquid equation Eq.~\eqref{eqLLCorrelationsZmag}, for finite inter-chain coupling $J_y$ and different magnetization values $m$, see Fig.~\ref{FigAppGS_corrFits}(a)--(d).
The results for $\tilde{K}$ obtained from these fits were used for Fig.~\ref{FigSix_correlations}(a) and Fig.~\ref{FigSix_correlations}(b) in the main text.

Absolute value of the correlations for zero magnetization $m = 0$, corresponds to $C^{z}_{AFM_{x}}(x)$, Eq.~\eqref{eqSzSzStagCorrelator} in the main text.
This correlator saturates to the finite value in the AFM stripe regime in the long distance limit, see Fig.~\ref{FigAppGS_corrFits}(e), and decays to zero in the disordered regime, see Fig.~\ref{FigAppGS_corrFits}(f).
Extracting the exact value where $\lim_{x \rightarrow \infty} (-1)^x \langle \Sh^{z}_{x_0, y_0} \Sh^{z}_{x_0+x, y_0} \rangle \neq 0$, is challenging, and we thus use the vale of $\tilde{K}$ to estimate the transition between AFM stripes and the disordered phase in the main text.

For finite magnetization we fit the absolute vale of the spin-spin correlations with Eq.~\eqref{eqStripeFct} from the main text, see Fig.~\ref{FigAppGS_corrFits}(g) and Fig.~\ref{FigAppGS_corrFits}(h).
The parameter $\varphi$ in the fitting function Eq.~\eqref{eqStripeFct} captures the period of the super-stripes, while the parameter $A$ reflects the amplitude of the super-stripes. 
In addition, parameters $B$ and $\alpha$ characterize the decay behavior outside the super-stripe regime, whereas the parameter $C$ accounts for any finite non-oscillatory contribution arising from the finite total magnetization $m$.
Furthermore, the parameter $\theta$ is included to correctly capture the phase of the super-stripe pattern, which may depend on both the number of stripes and the system size in the $x$-direction, for which open boundary conditions were employed.
The super-stripes are visible in Fig.~\ref{FigAppGS_corrFits}(g), where the wave pattern can be identified.
In the disordered regime, shown in Fig.~\ref{FigAppGS_corrFits}(h), the correlations decay much faster.

\begin{figure*}[t]
\centering
\epsfig{file=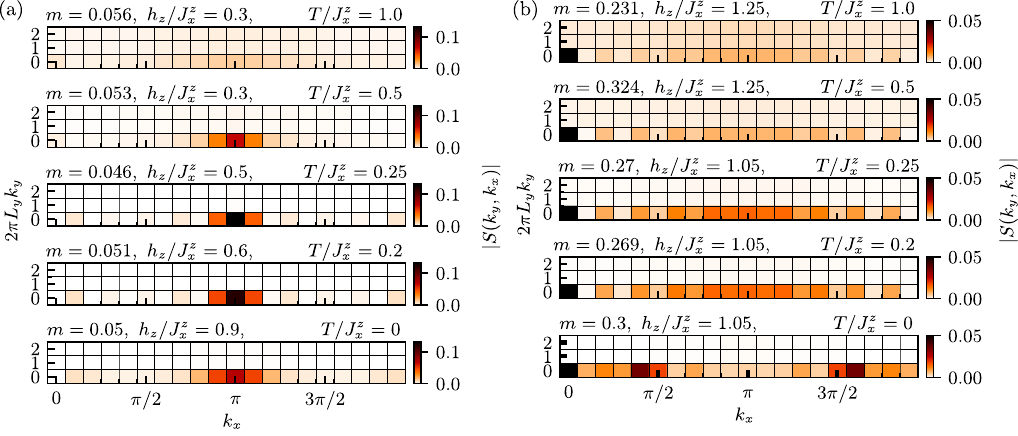}
\caption{Numerical results for the Fourier transformation Eq.~\eqref{eqFourier2DSpinSpin} of the spin-spin correlations for a system size of $L_x = 20$ and $L_y = 3$. (a) At lower magnetization $|m| \approx 0.05 $ peaks at $k_x = \pi$ remain well defined up to relatively high temperature $T \approx 0.5 J_x^{z}$, and broaden at high temperature $T \gtrsim J_x^{z}$. At such low magnetization and system size the peaks at $k_x = 1 \pm 2m$ in the ground state, $T = 0$, can not be resolved, and appear as a weak single peak at $k_x = \pi$. (b) At higher magnetization $|m| \approx 0.3$ the peaks at $k_x = \pi$ are relatively broad already for low temperatures $T \approx 0.2 J_{x}^{z}$ in comparison to other temperatures, which suggest an immediate transition to a chargon gas from the stripe phase at $T = 0$. The parameters are similar to those in Fig.~\ref{FigEight_mesonG}, i.e., $J^{\perp}_{x} / J^{z}_{x} = 0.4$ and $J_{y} / J^{z}_{x} = 0.8$ .
}
\label{FigAppFourierSpinSpinMagnetDiff}
\end{figure*}

\subsection{Finite temperature calculations}
The finite-temperature calculations are performed by employing the quantum purification scheme, where we enlarge our system by connecting an ancillary lattice site to every physical lattice site \cite{Zwolak2004, Feiguin2005, Feiguin2013, Nocera2016}.
We then construct the maximally entangled state $\ket{\psi(\beta = 0) }$, between every physically site and its corresponding ancilla site.
This is realized by calculating the ground state of an entangler Hamiltonian \cite{Nocera2016} using DMRG
\begin{equation}
    \H_e = - \sum_j \le \hat{S}^{+}_{p(j)} \hat{S}^{-}_{a(j)} + \hc \r.
    \label{eqEntangler}
\end{equation}
Here we defined $p(j)$ as the physical lattice sites and $a(j)$ as its corresponding ancilla lattice sites.
Such state then corresponds to the infinite temperature state $\beta = 1 / T = 0$.

We then perform imaginary time evolution of the system
\begin{equation}
    \ket{\psi (\beta)} = e^{\beta \H_t / 2} \ket{\psi(\beta = 0) } ,
\end{equation}
where the Hamiltonian $\H_t$ acts only on the physical lattice site.
We employ a Krlyov algorithm for the first few time steps which are followed by the time evolution using a $2$TDVP algorithm \cite{Kebric2023FiniteT, Paeckel2019}.
In order to obtain finite magnetization of the physical system we add the field term to the usual mixed-dimensional Hamiltonian
\begin{equation}
    \H_t = \H_{\rm XXZ} - h_z \sum_{j  = 1}^{LxLy} \Sh^{z}_{p(j)} ,
    \label{eqImagEvol}
\end{equation}
and we employ the U$(1)$ symmetry in the whole system of ancilla and physical sites.
We thus remain in the sector where $m = 0$ of the whole system.
The method however does not conserve the U$(1)$ symmetry separately in the ancilla or physical sites \cite{Nocera2016}, thus allowing us to obtain finite magnetization in the physical sites.
As a result, the values of magnetization $m$ at finite $T$ slightly differ from the exact target values because we had to perform imaginary time evolutions for a set of different field values $h_z$, and select the results where $m$ approached the target value best.
Further increasing the precision of the magnetization $m$ at finite temperature $T$ would require a substantial increase in the number of field values $h_z$, for which time evolutions are performed.
This would significantly increase the overall computational cost.
We believe that the current precision is already sufficient to capture the relevant physics, and further improvements in precision would not qualitatively affect the behavior of the system.

Any physical observable is then calculated by simply calculating the expectation value of the observable which only acts on the physical lattice site \cite{Nocera2016}
\begin{equation}
    \left \langle \mathcal{\hat{O}} \right \rangle =
    \frac{ \langle \psi(\beta) | \mathcal{\hat{O}} | \psi(\beta) \rangle} { \langle \psi(\beta) | \psi(\beta) \rangle } .
\end{equation}

\subsection{Fourier transformation of the spin-spin correlations at different magnetization \label{MoreFiniteTspinCorrelations}}

In this section we present further numerical results of the Fourier transformation of the spin-spin correlations, Eq.~\eqref{eq2DSpinSpin}.
We show the results for different magnetization in the same parameter regime as in Fig.~\ref{FigEight_mesonG}, which further support our schematic phase diagram sketched in Fig.~\ref{FigOne}(f).

In Fig.~\ref{FigAppFourierSpinSpinMagnetDiff}(a) we present the results for low magnetization $|m| \approx 0.05$, where the peaks around $k_x = \pi$ remain localized at high temperature $T \approx 0.5 J_{x}^{z}$, relative to the results seen for $|m| = 0.15$ in Fig.~\ref{FigEight_mesonG}.
This supports our claim that at low magnetization the underlying AFM order remains stable for higher temperature $T$, which can host the confined meson gas.
We note that peaks at $k_x = \pi (1 \pm 2m) $ in the ground state, $T = 0$, are difficult to distinguish, due to the relatively small system size.
Namely, our resolution is limited by $\Delta k_x = \frac{2 \pi}{L_x} $ and $\Delta k_y = \frac{2 \pi}{L_y}$, respectively.
This puts us at the limit when $|m| \approx 0.05$ for $L_x = 20$.
Due to the proximity of the peaks at $k_x = 0.9 \pi, 1.1 \pi$ a strong signal can be seen at $k_x = \pi$ instead.

At much higher magnetization, $|m| \approx 0.3$, peaks at $k_x = \pi$ are broad for any finite temperature in the range, which was within reach of our numerical calculations; see Fig.~\ref{FigAppFourierSpinSpinMagnetDiff}(b).
In the ground state, $T = 0$, we observe distinct peaks at $k_x = \pi (1 \pm 2m)$.
This suggest that the super-stripe order melts to a chargon gas already at low temperature, and we do not find a meson gas state in our numerical results.
The exact value of magnetization and temperature at which the transition occurs, remains an open question for further work and goes beyond the scope of this work.
However, from thermodynamic arguments we expect to find a stable super-stripe region at low temperature until thermal fluctuations destroy such order.
These arguments together with the results presented in the main text were used to schematically sketch the phase diagram presented in Fig.~\ref{FigOne}(f).

\begin{figure}[t]
\centering
\epsfig{file=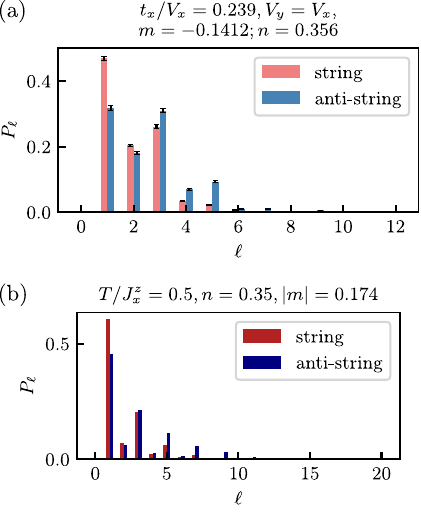}
\caption{String and anti-string length distributions. (a) Experimental results for tunneling amplitude $t_x / V_x = 0.239$, $V_y = V_x$, and magnetization $|m| = 0.141$. The string-length peak at $\ell = 1$ is slightly higher than the corresponding anti-string peak. We also observe a noticeable signal for strings and anti-strings of length $\ell = 3$.
(b) Numerical MPS results showing a similar trend to the experimental data, where the string-length peak at $\ell = 1$ is slightly higher than the anti-string peak. However, the peaks at $\ell = 3$ are smaller than those observed experimentally.
For the numerical results, the parameters used were $J_{x}^{\perp} / J_{x}^{z}= 0.4$, $J_{y} / J_{x}^{z} = 0.8$, which correspond to $t = 0.2V_{x}$ and $V_y = 0.8V_{x}$.
The data presented here are the same as the probability distributions shown in Fig.~5 of the main text.
}
\label{FigAppendixStringHistogram}
\end{figure}

\subsection{Additional results for string-length distributions}
Here we show additional data for the string and anti-string length distributions corresponding to the experimental and numerical results shown in Fig.~5 of the main text, where we present the probability distributions of the distance between partons.
In Fig.~\ref{FigAppendixStringHistogram}(a), we show the experimental results for $t_x / V_x = 0.239$, $V_y = V_x$, and magnetization $|m| = 0.141$, whereas Fig.~\ref{FigAppendixStringHistogram}(b) shows the MPS results.
In both cases, the string-length peaks at $\ell = 1$ are higher than the corresponding anti-string peaks.
However, the peaks for strings and anti-strings at $\ell = 3$ are more prominent in the experimental data than in the numerical results.
Overall, the string and anti-string length distributions are consistent with the corresponding probability distributions of the distance between partons shown in Fig.~5 of the main text.

\subsection{Parton spacing distributions \label{AppendixPartonSpacing}}
Here we provide details on calculating the parton distributions from snapshots.
In each snapshot we analyze the one-dimensional chains that form our mixD system, see Fig.~1(b) of the main text.
We sweep through the chains and record the distance $r$ between individual partons labeled, $a$ and $b$, respectively.
We note that partons (\Zt charges) correspond to the AFM domain walls in the mixD spin model Eq.~\eqref{eqMixedDXXZModel}.
This allows us to construct histograms and we plot the density distributions $p_{a,b}(r)$.

\begin{figure}[t]
\centering
\epsfig{file=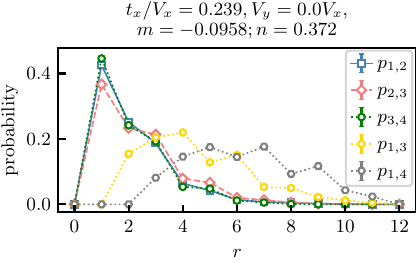}
\caption{Probability distribution of the distance between partons $p_{a,b}(r)$ obtained from the experimental data for $t_x / V_x = 0.239$ and $V_y = 0$. Distributions $p_{1,2}, p_{2,3}, p_{3,4}$ are qualitatively similar as expected for the deconfined regime when the chains are decoupled.
}
\label{FigDeconfinedExperiment}
\end{figure}

As discussed in the main text, we can not easily determine the identity of strings and anti-strings.
Hence, we use the value of the staggered magnetization to determine whether our first parton in the system has a string attached from the left or from the right.
The same procedure is used to construct the string and anti-string length histograms, see Section~\ref{SectionV_strings} of the main text.
We can use this procedure as strings are much shorter than anti-strings in the confined meson gas regime, which results in a finite value of staggered magnetization, defined as 
\begin{equation}
    m_s^{y} =  \frac{1}{L_x} \sum_{x} (-1)^j \langle \psi | \Sh^{z}_{x, y \pm 1} | \psi \rangle.
\end{equation}
The sign of the staggered magnetization can be used to determine whether our system starts or ends with a string.
In the case where the staggered magnetization is positive $m_s^{y} > 0$, we switch the identity of partons as $2s-1 \rightarrow 2s$ and $2s \rightarrow 2s + 1$.
In other words the distributions become $p_{a, b}(r) \rightarrow p_{a+1, b+1}(r)$.
We perform this for our experimental as well as numerical snapshots.
This procedure successfully determines the confined phase in our system as we observe that $p_{1,2}(r=1) \approx p_{3,4}(r=1) > p_{2,3}(r=1) $, in the regime where we expected the meson gas.
In the deconfined, uncoupled regime $V_y = J_y=0$, these distributions should remain similar, $p_{1,2}(r=1) \approx p_{3,4}(r=1) \approx p_{2,3}(r=1) $, as partons do not not confine into mesons.
We confirm this for our experimental results in Fig.~\ref{FigDeconfinedExperiment} for parameters $t_x / V_x = 0.239$, $V_y = 0$, where $p_{1,2}(r) \approx p_{3,4}(r) \approx p_{2,3}(r) $.
This is in contrast to the observed confined signatures in Fig.~\ref{FigFive_MesonGasDistances}(a) in the main text in the coupled regime, $V_y = V_x$, for the same tunneling rate, and comparable magnetization as in the decoupled regime in Fig.~\ref{FigDeconfinedExperiment}.
This shows that our method can be used to correctly distinguish between confined and deconfined regimes.

Since our system does not conserve the total parton number $N^{a}$, some snapshots may contain less partons that we need to analyze our distributions $p_{a,b}(r)$, i.e., it can happen that in some snapshots $N^a < (b+1)$, where $b>a$.
This is because our system is finite and we could reach the end of our chain before reaching parton $b$.
Note that due to the possible switch of the string and anti-string identity we wrote $(b+1)$ instead of $b$.
This can results in distributions $p_{a,b}(r)$ to have finite weight at $r<|a-b|$,  which is nonphysical.
This is due to the fact that in such case we have reached the end of the chain but did not encounter the $b$-th parton.
We overcome this problem by simply ignoring nonphysical lengths $r<|a-b|$, for distribution $p_{a,b}$, while we retain all the physical distributions $p_{m,n}$ for our analysis.
This results in slightly enhanced distributions weight for low $r$ for $p_{1,3}$ and $p_{1,4}$, as we reach the end of the system before reaching parton $b= 3$ or $b = 4$, respectively. 
However, simply ignoring entire snapshots where $N^a < b$ (or $N^a < (b+1)$, depending on the staggered magnetization) would not give us correct distributions.

\begin{figure}[t]
\centering
\epsfig{file=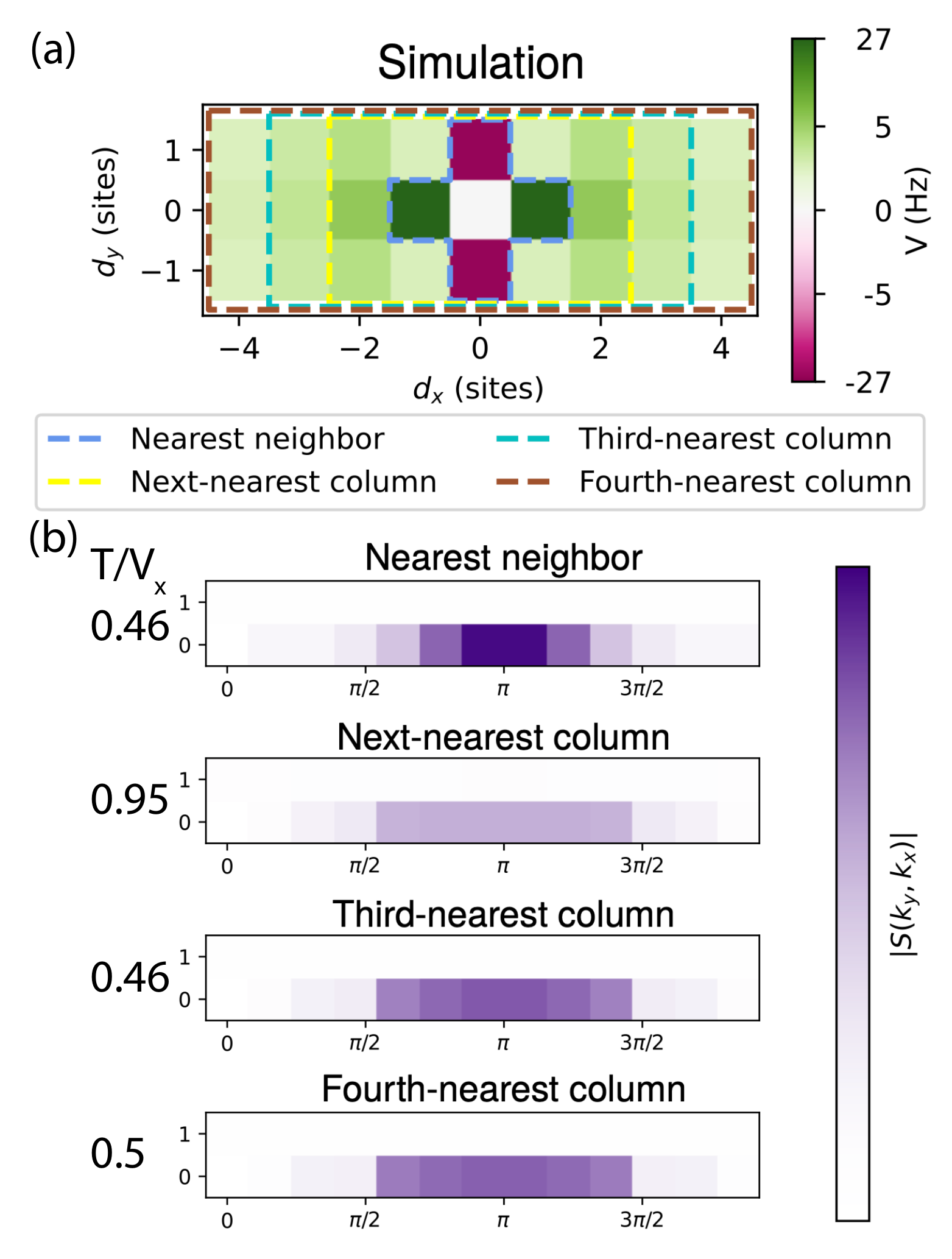, width=.45\textwidth}
\caption{Dipolar interaction stabilizes super-stripes at finite temperature and reduces the presence of meson gas peak in the structure factor for non-zero magnetization. (a) We simulate the finite temperature states by truncating to different interaction ranges using Exact Diagonalization. The range of dipolar tails included in the simulation is gradually increased. (b) By truncating the dipolar interaction in the numerical simulation to only nearest neighbor interactions, we reproduce results shown in Fig.~\ref{FigEight_mesonG} at the temperature of $T=0.46V_x$. By adding the next-nearest column interactions the meson gas feature dramatically decreases and the most prominent peak happens at $T=0.95V_x$. Increasing the truncation to the third nearest column slightly promotes the meson gas feature at $T=0.46V_x$. But adding the fourth nearest column washes out the feature a bit more and the most prominent peak happens at $T=0.5V_x$. These results demonstrate the frustration induced by long-range dipolar interaction on the meson gas phase. The system size was $L_x = 9$ and $L_y = 2$.
}
\label{FigED_dipolar}
\end{figure}

\section{Experimental details \label{ExperimentalDetailsAppendix}}

\subsection{Estimate of chemical potential spatial disorder in experiment}
We measured the peak-to-peak disorder of our vertical lattice and tight-spacing 2D lattices to be roughly 5 Hz \cite{Su2023} within our region of interest of two hundred sites. The harmonic confinement of our lattices is compensated with optical potentials generated by a Digital Micro-Mirro Device. This compensation can introduce up to another 5Hz of estimated disorder, especially on the edge of the region of interest. We measured the peak-to-peak disorder of our tunable spacing super lattice to be roughly 5 Hz during the initial state preparation, where the lattice depth is approximately 200 Hz. The three sources of disorders are uncorrelated, so the total disorder is less than 10 Hz when the superlattice is turned on for state preparation. In the tunneling ramp later, the disorder is estimated to be 7 Hz peak to peak. In comparison, the dipolar interaction $V_x$ is roughly 25 Hz, which is more than a factor of 3 larger.

\subsection{Adiabaticity of tunneling ramp}
We choose the tunneling ramp duration based on two considerations. On one hand, we try to ramp as fast as possible, so that heating due to technical noise (off-resonance scattering, intensity noise, etc.) is minimized. On the other hand, especially when ramping across quantum phase transitions, we prefer to go slower than the many-body energy gap so that the majority of the many-body wave-function remains in the ground state after the ramp. We have chosen the ramp duration to be roughly 100 ms for the data taken in this paper. We use exact diagonalization to simulate the ramp of a small system (3 rows, 6 sites per row, periodic boundary conditions). We checked that ground state fidelity is around 95\% for all ramp parameters used in this paper. Since our system is much larger, the ground state fidelity will be lower, but this qualitatively demonstrates that we are relatively close to being adiabatic in our ramps.

\subsection{Effects of dipolar tails on the meson gas}
The inclusion of dipolar interaction beyond the nearest neighbor modifies the phase diagram for $m\neq0$ at finite temperature by enlarging the region of super-stripes and reducing the region of meson gas. With only nearest-neighbor interactions and in the limit of negligible tunneling $t$, a column of atoms mobilizing to the neighboring column is not penalized by an increase of potential energy. Therefore, the super-stripe phase is relatively fragile. However, when long-range dipolar interaction is included, the super-stripe is stabilized since moving a column of atoms becomes energetically unfavorable. We perform exact diagonalization simulations for small system sizes of 2 by 9 sites to include the long-range dipolar interactions shown in Fig.~\ref{FigED_dipolar}, where the sharpness of the peak at commensurate $\mathbf{k}=(\pi, 0)$, associated with the meson gas, is washed out as the dipolar interactions from the next-nearest columns are included.


%


\end{document}